\def\Granada{Instituto Carlos I de F\'\i sica Te\'orica y Computacional, 
Facultad de Ciencias, Universidad de Granada, Campus de Fuentenueva,
Granada 18002, Spain}

\def\Valencia{IFIC, Centro Mixto Universidad de Valencia-CSIC, Burjasot
              46100-Valencia,Spain.}

\def\INFN{Dipartimento di Scienze Fisiche and INFN, Sezione di Napoli,
      Mostra d'Oltremare, Pad. 19, 80125, Napoli, Italia.}

\def\Comision{Work partially supported by the Direcci\'on General de Ciencia y
              Tecnolog\'\i a.}

\def\medio{\frac{1}{2}}
\def\nn{\nonumber}

\def\PHO{{\cal P}^{HO}}
\def\ni{\noindent}
\def\cL{{\cal L}}

\def\GC{{\cal G}_C}
\def\XL{\tilde{X}^L}
\def\XR{\tilde{X}^R}
\def\EG{$\tilde{G}_{(m)}$}
\def\Gt{$\tilde{G}\,$}
\def\Gtm{\tilde{G}}
\def\z{\zeta}

\def\x{\vec{x}}
\def\q{\vec{q}}

\def\a{\vec{a}}
\def\v{\vec{v}}
\def\w{\omega}
\def\r{\vec{\rho}}
\def\c{\vec{c}}
\def\[{\left[}
\def\]{\right]}
\def\ba{\begin{array}}
\def\ea{\end{array}}
\def\be{\begin{equation}}
\def\ee{\end{equation}}
\def\bea{\begin{eqnarray}}
\def\eea{\end{eqnarray}}
\def\H{{\cal H}}

\def\hx{\tilde{x}}
\def\hy{\tilde{y}}
\def\htau{\tilde{\tau}}
\def\hPsi{\widetilde{\Psi}}
\def\hchi{\widetilde{\chi}}

\newcommand{\parcial}[1]{ \frac{\partial}{\partial #1} }

\documentstyle[12pt]{article}

\title{ QUANTIZATION ON A LIE GROUP: HIGHER-ORDER POLARIZATIONS 
           \thanks{\Comision} }

\author{ V. Aldaya$^{1,2}$, J. Guerrero$^{1,3}$,  
         and G. Marmo$^{3}$}         

\begin{document}


\setcounter{page}{0}

\footnotetext[1]{\Granada} \footnotetext[2]{\Valencia}
\footnotetext[3]{\INFN}

\maketitle

\section{Introduction}
Much effort has been devoted to the geometrization of Quantum Mechanics 
during the second half of this Century in an attempt to emulate Classical
Mechanics and Classical Gravity at mathematical beauty and, why not, to 
better understand Quantum Gravity. We wish to report on one particular 
line of this task, which lies mostly on symmetry grounds and has been
developed in the last years trying to accomodate modern aspects of 
Quantum Mechanics such as global quantization of systems with non-trivial
topology, in particular systems suffering from topological anomalies, 
and accounting for more general obstructions to the basic rules of local 
quantization, to be referred to as algebraic anomalies, directly attached
to the well-known no-go theorems \cite{Hove} of the original, Standard Quantum
Mechanics. This Group Approach to Quantization (GAQ) \cite{23}, in some respect generalizes 
Geometric Quantization (GQ), originally developed by Kirillov, Kostant and 
Souriau \cite{Kirillov,Kostant,Souriau} 
both as a method of quantization and as a group representation technique, 
and the more specific representation algorithm of Borel-Weyl-Bott (B-W-B) 
\cite{Preisler}, 
which essentially applies to finite-dimensional semisimple groups.

The idea of turning groups into basic building blocks for the geometric formulation
of Physics is simply the natural result of pushing ahead the old
usage of imposing the compatibility of observer in the same way
Differential Geometry itself considers admissibility of local chart.
The requirement of a definite structure in the set of observers, or atlas,
delimites seriously the nature of physical laws in that they must be formulated 
in terms of say $GL(n,R)$-tensors, although this requirement is not
restrictive enough so as to actually ``predict" dynamical laws. However, the condition of 
having defined an associative composition law in a set of ``active" 
transformations of a physical system really predicts in many cases its 
dynamics, and can accordingly be considered as a basic postulate.

In the particular case of Quantum Mechanics the group structure plays 
a preponderant technical role also because, after all, the quantization 
map has always been viewed as a representation, in the Lie algebra sense, of 
definite subalgebras of the general Poisson algebra defined on classical 
phase space. In this respect, the group manifold along with canonical structures
on it will constitute a powerful tool in the practical construction 
of the quantum representation, mainly due to the existence of two natural,
mutually (commuting left and right) group actions. In fact, one of which can 
be used to compatibly reduce the regular representation (or some generalization 
of it) given by the other.
  
However, groups entering the basic postulates of (Quantum) Physics mostly 
wear a specific topological and algebraic structure which
goes around the notion of ``extension by $U(1)$" of both the ``classical"
symmetry group and the classical phase space, and we wish to dwell a bit on the
necessity of this extensions.

To help to visualize how the paper is organized, we give a list of contents:

\begin{itemize}
\item Introduction
      \begin{itemize}
      \item Why $S^1$-extended phase space?
      \item Why central extensions of classical symmetries?
      \end{itemize}
\item Central extension \Gt of a group $G$
      \begin{itemize}
      \item Group cohomology
      \item Cohomology and contractions: Pseudo-cohomology
      \item Principal bundle with connection $(\Gtm,\Theta)$
      \end{itemize}
\item Group Approach to Quantization
      \begin{itemize}
      \item $U(1)$-quantization
      \item Non-horizontal polarizations
      \end{itemize}
\item Simple examples
      \begin{itemize}
      \item The abelian group $R^{k}$ 
      \item The semisimple group $SU(2)$
      \end{itemize}
\item Algebraic anomalies
      \begin{itemize}
      \item Higher-order polarizations
      \item The Schr\"odinger group and Quantum Optics
      \item The Virasoro group and String Theory
      \end{itemize}
\end{itemize}


\subsection{Why $S^1$-extended phase space?}

\ni In a naive attempt to quantize geometricaly a classical phase space $M$, 
i.e. to represent irreducibly the Poisson algebra in terms of unitary operators
acting on some Hilbert space naturally attached to the manifold $M$, we are
primarily aimed to consider the Hamiltonian vector fields $X_f$ associated 
with the function $f$ as a ``quantum" operator $\hat{f}$ acting on 
$C^\infty(M)$ by derivation.

In practice, and in the simplest, though rather general, case we start from 
the space $R\times R^3\times R^3$ and construct the Poincar\'e-Cartan form 
associated with a Hamiltonian function $H$ \cite{Malliavin} and its 
differential, a presymplectic two-form:

\bea 
\Theta_{PC}&\equiv& p_idx^i-Hdt \\
\Omega &\equiv&d\Theta_{PC}=dp_i\wedge dx^i-\frac{\partial H}{\partial x^i}
    dx^i\wedge dt-\frac{\partial H}{\partial p_i}dp_i\wedge dt\,,\nn
\eea

\ni $\Omega$ has a one-dimensional kernel generated by $X_H\in Ker\, d\Theta_{PC}$
such that $dt(X_H)=1$,

\be
X_H=\frac{\partial \;}{\partial t}+\frac{\partial H}{\partial p_i}\frac{\partial \;}
     {\partial x^i}-\frac{\partial H}{\partial x^i}\frac{\partial \;}{\partial p_i}\,,
\ee

\ni and the associated equations of motion are the Hamilton equations:

\bea
\frac{dt}{d\tau}&=&1\nn\\
\frac{dx^i}{d\tau}&=&\frac{\partial H}{\partial p_i}\\
\frac{dp_i}{d\tau}&=&-\frac{\partial H}{\partial x^i}\,.\nn
\eea

The vector field $X_H$, if it is complete, defines a one-parameter group $\phi_{\tau}$ which 
divides the space of movements $R\times R^3\times R^3$ into classes, $M\equiv \{R\times R^3\times 
R^3\}/\phi_{\tau}$ constitutes the symplectic phase space of the system characterized 
by the Hamiltonian $H$; the symplectic form obtains by the projection of $\Omega$. 
The change of variables under which the equations of motion on the quotient
become trivial is the Hamilton-Jacobi transformation. As a trivial example let us
consider the case $H=\frac{p^2}{2m}$ corresponding to a free particle of mass $m$.
In this case the equations for $\phi_{\tau}$ and their inverses are:

\be
\left\{\ba{l}x^i=\frac{P^i}{m}\tau+K^i\\ p_i=P_i\\ t=\tau \ea \right.
\Longleftrightarrow \left\{\ba{l}K^i=x^i-\frac{p^i}{m}t\\ P_i=p_i\\ 
   \tau=t\,, \ea \right.
\ee

\ni where the constants of motion $K^i,P_j$ parametrize the solution 
manifold $M$. $\Theta_{PC}$ goes to the quotient except for a total differential:

\bea 
\Theta_{PC}&\rightarrow &P_idK^i+d(\frac{P^2}{2m}\tau)\\
\omega &=&dP_i\wedge dK^i\,.\nn
\eea 

The symplectic form \cite{Godbillon,Abraham} is an anti-symmetric ``metric" and 
defines an isomorphism
$\omega^\flat:{\cal X}(M)\leftrightarrow \Lambda^1(M)$ between the module 
of vector fields on $M$ and that of one-forms on $M$,

\be
X\in{\cal X}(M)\longmapsto\omega(X,\cdot)\equiv i_X\omega\in\Lambda^1(M)\,,
\ee

\ni associating a bracket $\{,\}$ on $\Lambda^1(M)$ with the Lie bracket of 
vector fields. In particular, given functions $f,g\in C^\infty(M)$, their 
differentials are associated with Hamiltonian vector fields $X_f,X_g$. This permits
the definition of a Poisson bracket between functions, rather than one-forms,
but this time the correspondence $\{,\}\rightarrow [,]$:

\be
\{,\}:f,g\longmapsto\{f,g\}\,\,/\,\,d\{f,g\}=-i_{\[X_f,X_g\]}\omega\,,
\ee

\ni is no longer an isomorphism because constant functions have trivial
Hamiltonian vector fields. In particular, as regard the example $H=\frac{p^2}{2m}$,
and considering $K^i,P_j$ as the basic coordinates for $M$, we find:

\be
\{K^i,P_j\}=\delta^i_j\cdot 1\longmapsto \[X_{K^i},X_{P_j}\]=0
\ee

\ni i.e. a Lie algebra homomorphism whose kernel is the central subalgebra of
constant functions, $R$, generated by $1$.

We should remark that the correspondence between functions and vector
fields can be kept at the movement-space level provided that we restrict ourselves
to only those functions which are invariant under the action of $X_H$, i.e. pull-backs 
of functions on $M$.

The existence of a non-trivial kernel in the correspondence between
functions and Hamiltonian vector fields is an essential breakdown to
the naive geometric approach to quantization $\hat{}:f\mapsto \hat{f}\equiv
X_f$ which would associate the trivial operator to any constant. The 
simplest way of avoiding this problem will consist in enlarging phase space 
(and/or movement space) with one extra variable providing one extra component to $X_f$, and 
generalizing accordingly the equation $i_{X_f}d\Theta_{PC}=-df$ so as to get a 
non-trivial new component even though $f$ is a constant. On a ``quantum 
manifold" $P$ locally isomorphic to $M\times S^1$ with ``connection form" 
$\Theta$ such that the curvature two-form $d\Theta$ induces on $M$ a 
2-form equal to the one induced by $\Theta_{PC}$, the 
equation above can be replaced by the set of equations \cite{Souriau}:
\bea
i_{\tilde{X}_f}d\Theta &=& -df\nn\\
i_{\tilde{X}_f}\Theta &=& f \label{ixdteta}
\eea

\ni generalizing this way the quantization map which now reads (except perhaps
for a minus sign) 

\be
\hat{f}\equiv i\tilde{X}_f
\ee 

\ni Locally, we can write $\Theta=\Theta_{PC}+\frac{dz}{iz},\;z\in S^1$ and then
$\tilde{X}_f=X_f+(f-\Theta(X_f))iz\frac{\partial \;}{\partial z}$, and we 
immediately see that (\ref{ixdteta}) have unique solution associating the
fundamental (vertical) vector field $\Xi\equiv iz\frac{\partial \;}{\partial z}$ with
the unity of $R$.

The quantization map $\hat{}$ is now an isomorphism between the Poisson 
algebra on $M$ and the Lie subalgebra of vector fields on $P$ that are solutions
to (\ref{ixdteta}). For the basic functions we have:

\be
\{K^i,P_j\}=\delta^i_j\cdot 1 \longleftrightarrow \[\tilde{X}_{K^i},
      \tilde{X}_{P_j}\]=\delta^i_j\cdot \Xi
\ee

The space of wave functions $\Psi$ is constituted by the complex-valued functions on $P$
which satisfy the $U(1)$-equivariance condition, turning $\Psi$ into a section
of an associate bundle of the principal bundle $P\rightarrow M$ \cite{Nomizu}:

\be 
\Xi\Psi=i\Psi \longleftrightarrow \Psi(K,P,z)=z\Psi(K,P)\,,
\ee

\ni on which $\tilde{X}_f$ act defining the ``quantum" operators.

Unfortunately, the quantization map $\hat{}$ is faithful but not irreducible
as a representation of the Lie algebra of classical functions. At this 
prequantization level, we are only able to
reproduce the Bohr-Sommerfeld-Wilson quantization rules \cite{Woodhouse}. We 
know that this
representation is reducible because of the existence of non-trivial operators
commuting with the basic quantum generators $\hat{K}^i,\hat{P}_i$. In fact, 
thinking of the simplest case, that of the free particle for example, and adopting
for $\Theta$ the local expression,

\be 
\Theta=P_idK^i+\frac{dz}{iz} \,,
\ee

\ni we get the following basic operators: 

\bea
\hat{K}^i &=&i\frac{\partial \; }{\partial P_i}+K^i \nn\\
\hat{P}_j &=&-i\frac{\partial \; }{\partial K^j}\,,
\eea

\ni and it is clear that the operator $\frac{\partial \; }{\partial P_i}$ 
does commute with them.

True quantization requires that we determine a subspace of complex-valued functions on which
our quantum generators act irreducibly.
We must then impose a maximal set of mutually compatible 
conditions in the form $\tilde{X}\Psi=0$, for $\tilde{X}$ in some maximal
subalgebra called polarization. For instance, in the example above the 
operator $\frac{\partial \;}{\partial P_i}$ would become trivial had we selected our 
subspace by imposing 
the polarization condition $\frac{\partial \;}{\partial P_i}\Psi=0$, i. e.
$\Psi\neq\Psi(P)$. Finding a polarization, however, is not an easy task in 
general, the reason being that two polarization conditions $\hat{a}\Psi=0,
\hat{b}\Psi=0$ would be inconsistent, for instance if $[\hat{a},\hat{b}]=\hat{1}$. 
In addition,
once a certain polarization has been imposed, the set of physical operators
that preserve the polarization is severely restricted.


\subsection{Why central extensions of classical symmetries?}

\ni The close connection between central extensions and quantization is best illustrated
by the simple case of the free particle and its {\it semi-invariance} 
properties \cite{Levy-Leblond,Sudarshan}, which has its origin in the specific algebraic 
structure of the group. Let us dwell on 
the semi-invariance of the free particle in both classical and quantum versions 
in $1+1$ dimensions. The most relevant feature is the lack of exact 
invariance of the Lagrangian
 
\begin{equation}
\cL=\frac{1}{2}m\dot{x}^2
\end{equation}

\noindent under the Galilean transformations:

\begin{eqnarray}
x'&=&x+A+Vt \nn \\ 
t'&=&t+B \,.     \label{Galileo}
\end{eqnarray}

\noindent In fact, the transformed Lagrangian acquires an extra term:

\begin{eqnarray}
\cL'&=&\frac{1}{2}m(\dot{x}+V)^2=\frac{1}{2}m\dot{x}^2+\frac{1}{2}mV^2+
        m\dot{x}V  \nonumber \\
&=&\cL+\frac{d}{dt}(\frac{1}{2}mV^2t+mVx) \,,
\end{eqnarray}

\noindent which is a non-zero total derivative not affecting the extremals of the 
corresponding action. The same phenomenon appears at the infinitesimal level.
The infinitesimal generators associated with the finite transformations 
(\ref{Galileo}) are obtained by deriving the transformed coordinates with
respect to each one of the group parameters at the identity $e$ of the group:

\begin{eqnarray}
X_B&=&\frac{\partial}{\partial t} \nn \\
X_A&=&\frac{\partial}{\partial x} \\
X_V&=&t\frac{\partial}{\partial x}\,. \nn
\end{eqnarray}

\noindent When we compute the Lie derivative of $\cL$ with respect to
these generators, i.e. the directional derivative of $\cL$ along trajectories of each 
generator of the group, we find

\begin{eqnarray}
X_B.\cL&=&0 \nn \\
X_A.\cL&=&0 \\
X_V.\cL&=&\frac{d\,}{dt}(mx) \neq 0\,, \nn
\end{eqnarray}

\ni i.e. the semi-invariance of $\cL$ under the {\it boosts} comes out again.
In a complete equivalent way, the Poincar\'e-Cartan
form (see, e.g. \cite{RNC}), 

\be
\Theta_{PC}\equiv pdx-Hdt=\frac{\partial \cL}
{\partial \dot{x}}dx-(p\dot{x}-\cL)dt=\frac{\partial \cL}{\partial \dot{x}}
(dx-\dot{x}dt)+\cL dt \,,
\ee

\ni also gives a total differential when the Lie derivative along boots is computed: 

\be
L_{X_V}\Theta_{PC}=d(mx)\,.
\ee

The quantum Galilean particle suffers from the same pathology although it shows 
up in a rather different manner. Let us consider the Schr\"odinger equation:

\begin{equation}
i\hbar\frac{\partial \,}{\partial t}\Psi=-\frac{\hbar^2}{2m}\nabla^2\Psi\,, \label{Schr}
\end{equation}

\ni and apply the Galilei transformations (\ref{Galileo}) to it. The equation
(\ref{Schr}) acquires an extra term,

\begin{equation}
i\hbar\frac{\partial \,}{\partial t'}\Psi+i\hbar V\frac{\partial \Psi}{\partial x'}
  =-\frac{\hbar^2}{2m}{\nabla'}^2\Psi \,, \label{Schrp}
\end{equation}

\ni which can be compensated by transforming the wave function. Allowing 
for a non-trivial phase factor in front of the transformed wave function of 
the form

\begin{equation}
\Psi'=e^{\frac{im}{\hbar}(Vx+\frac{1}{2}V^2t)}\Psi\,, \label{Psip}
\end{equation}

\ni the Schr\"odinger equation becomes covariant, i.e.

\begin{equation}
i\hbar\frac{\partial \,}{\partial t'}\Psi'=-\frac{\hbar^2}{2m}{\nabla'}^2\Psi' \,. 
\end{equation}

The need for a transformation like (\ref{Psip}) accompanying the space-time
transformation (\ref{Galileo}) to accomplish full invariance strongly suggests
the adoption of a central extension of the Galilei group as the basic symmetry
for the free particle \cite{Bargmann}. The $\hbar$ constant is required to 
keep the exponent in (\ref{Psip}) dimensionless (of course the particular value of $\hbar$ 
is not fixed by this dimensional argument). At this point it should be 
remarked that the strict invariance of the classical Lagrangian could also be achieved by 
a central extension of the Galilei group by the additive group of 
the real line (a local version of the U(1) central extension of quantum 
symmetry). In that case the transformation analogous to (\ref{Psip}) should 
be

\begin{equation}
S'=S+m(Vx+\frac{1}{2}V^2t)\,,
\end{equation}

\ni where S is the principal Hamilton function, which wears the dimensions 
of an action, thus avoiding the need for the fundamental constant 
$\hbar$ \cite{23}. 

The successive composition of two transformations in the extended Galilei 
group $\tilde{G}_{(m)}$ immediately leads to the group law:

\begin{eqnarray}
B''&=&B'+B \nn \\
A''&=&A'+A+V'B \nn \\
V''&=&V'+V \\
\zeta''&=&\zeta'\zeta e^{\frac{im}{\hbar}[A'V+B(V'V+\frac{1}{2}{V'}^2)]} \,. \nn
\end{eqnarray}

\ni  The virtue of the central extension is that of making non-trivial the 
commutator between the generators associated with translations and boosts: 
$[\tilde{X}_A,\tilde{X}_V]=
m\Xi$, just mimicking the Poisson bracket between $p$ and $x$
($K$ and $P$ to be precise) provided that we impose on the wave function the 
$U(1)$-equivariance condition $\Xi\Psi=i\Psi$. This commutator, 
responsible of the (Lie-algebra) extension, will dictate in this and more 
general {\it quantization groups} the dynamical character (coordinate-momentum 
character) of the group parameters \cite{23}.

\section{Central extension \Gt of a group $G$}

The case of the 1+1 extended Galilei group \EG is an example of a central
extension of a group $G$ by $U(1)$. In the general case the group manifold
of the central extension \Gt is the direct product $G\times U(1)$, a fact which 
permits the parametrization of the group by the pair $\tilde{g}=(g,\zeta),\,\, g\in G,\,\,
\zeta\in U(1)$. However, the composition law does not correspond to a
direct product of groups. Instead, it is written in the form:

\bea
g''&=&g'*g \nn \\
\zeta''&=&\zeta'\zeta e^{i\xi(g',g)}\,,\label{central}
\eea

\ni where $\xi$ is a function with arguments in $G$. In this way, the identification
of elements $g\in G$ with $(g,1)\in \Gtm$ will not provide us with a subgroup of \Gt.
The elements $(e,\zeta)$,
where $e$ is the identity in $G$, commute with the whole group, i.e. 
are central. For (\ref{central}) to be a group law the function $\xi$ must 
satisfy

\begin{eqnarray}
\xi(g_1,g_2) + \xi(g_1*g_2,g_3) &=& \xi(g_1,g_2*g_3) + \xi(g_2,g_3) \nn\\
\xi(e,g)=0 \, &,& \,\xi(g,e)=0 \,.  \label{cociclo}
\end{eqnarray}

However, if there exists a function $\lambda:G\rightarrow R$ such that

\bea
\xi(g',g)&=&\lambda(g'*g)-\lambda(g')-\lambda(g)\nn \\
\lambda(e)&=&0\,,\label{coborde} 
\eea

\ni we can always resort to a change of variables

\bea
\bar{g}&=&g \\
\bar{\zeta}&=&\zeta e^{-i\lambda(g)}\,, \nn
\eea

\ni to take (\ref{central}) to the direct product law $\bar{g}''=
\bar{g}'*\bar{g},\,\bar{\zeta}''=\bar{\zeta}'\bar{\zeta}$.


\subsection{Group cohomology}

The construction above suggests a cohomological structure.
A function $\xi:G\times G\rightarrow R$ satisfying (\ref{cociclo}) is called a
{\it two-cocycle},
and we say that two two-cocycles are cohomologous if they differ by a coboundary, 
i.e. a two-cocycle of the form (\ref{coborde}). The function $\lambda$ is said to be
the generating function of the coboundary. Central extensions are then
characterized by the quotient $H^2(G,R)$ of the set, abelian group indeed, of
two-cocycles, modulo the subgroup of coboundaries, and is called the $2^{nd}$
cohomology group of $G$. An equivalence class in $H^2(G,R)$ will be 
denoted by $[[\xi]]$. The general theory of central extensions was 
formulated in \cite{Bargmann} (see also \cite{Kurosh,Michel}), and a 
more detailled study of Lie group and Lie algebra cohomology can be 
found in \cite{23}, which includes the connection to symplectic 
cohomology.


\subsection{Cohomology and contractions: Pseudo-cohomology}

\ni By {\it pseudo-cohomology} we generally mean a cohomology subclass $[\xi]\in[[\xi]]$
that can be distinguished inside the trivial cohomology, i.e. selected from 
the coboundaries, according to some additional structure associated with the 
group $G$. Pseudo-cohomology phenomenon was firstly studied by \cite{Saletan}
and then applied to relativistic quantization in Ref. \cite{Pseudo}. The physical
origin of pseudo-cohomology can be easily stated as follows. We start from a 
given group $G$ for which we know a central extension \Gt associated with a
two-cocycle $\xi_{cob}$ generated by a function $\lambda$ on $G$. Suppose
there is a well-defined contraction limit of the group \Gt giving $\Gtm_c$, in 
the sense of In\"on\"u and Wigner \cite{InonuWigner,Inonu}. This means, 
in particular, that the two-cocycle $\xi_{cob}$ is well-behaved under the limit process.
But it could well happen that the generating function $\lambda$, however, be
ill-defined (infinite) in this limit. Then, the contracted two-cocycle is no longer
a coboundary since there is no $\lambda_c$ generating it. The Lie algebra 
structure constant associated with a pseudo-extension, i.e. a central extension 
characterized by a pseudo-cocyle, really differs from those of the trivial product, 
a fact which requires the non-triviality of the gradient of $\lambda$ at the 
identity of the group $G$. In other words, pseudo-cocycles are generated by functions
which are, locally, linear functions. For finite-dimensional semisimple 
groups, for which the Whitehead Lemma applies \cite{Jacobson},
pseudo-cohomology is absolutely relevant and is also related to the $\check{C}$ech  cohomology 
of the generalized Hopf fibration by the Cartan subgroup $H$, $G\rightarrow G/H$,
\cite{Formal}. In the general case, including infinite-dimensional semisimple Lie groups, 
for which the Whitehead lemma does not apply, the group law for 
$\tilde{G}$ will contain two-cocycles as well as pseudo-cocycles 
(see \cite{Virasoro,Formal,Anomalias}). The simplest
physical example of a quantum symmetry including such an extension is that
of the free non-relativistic particle with spin; the Galilei group must
be extended by a true two-cocycle to describe the canonical commutation relations
between q's and p's as well as by a pseudo-cocycle associated with the 
Cartan subgroup of $SU(2)$, to account for the spin degree of freedom 
\cite{Position}.
 

\subsection{Principal bundle with connection $(\Gtm,\Theta)$}

\ni Central extensions constitute the simplest, but sufficiently general, examples
of joint structures which are basic in the quantization process. In fact, a central 
extension by $U(1)$ is in turn a Lie group and a $U(1)$ principal bundle on which
a connection 1-form can be naturally defined. They are trivial as mere 
fibre bundle, but not as principal bundles with connection. Of 
course, more general groups  bearing a non-trivial principal bundle
structure, with structure group $H$ can be considered, but an appropriate 
pseudo-extension by a two-cocycle generated by a locally linear function of the 
fibre parameters provides
exactly the same result \cite{Formal}.

The connection 1-form $\tilde{\theta}\equiv \Theta$ on $\tilde{G}$ will be 
selected among the components of the left-invariant, Lie algebra-valued, 
Maurer-Cartan 1-form, in such a way that $\Theta(\Xi)=1$ and 
$L_{\Xi}\Theta=0$, where $\Xi$ is the vertical 
(or fundamental) vector field. $\Theta$ will play the role of a 
connection 1-form associated with the $U(1)$-bundle structure. 

If \Gt is the topological product of $G$ and $U(1)$ (as 
will be always the case in all the examples we are considering), the group
law for \Gt can be written as:
\be
(g',\z')*(g,\z)=(g'*g,\z'\z e^{i\xi(g',g)})\, .
\ee

Considering a set of local canonical coordinates at the identity 
$\{g^i,\,\, i=1,\ldots, {\rm dim} G\}$ in $G$, the group law can be given by the
set of functions  $g''{}^i=g''{}^i(g'{}^j,g^k,\,\, j,k=1,\ldots,{\rm dim} G)$.
We introduce the sets of left- and right-invariant vector fields of \Gt
associated with the set of canonical coordinates $\{g^i\}$ as those which
 are written as $\parcial{g^i},\,\,i=1,\ldots,{\rm dim} G$ at the identity, 
that is:
\be \XL_{g^i}(\tilde{g}) = \tilde{L}^T_{\tilde{g}} \parcial{g^i}\,,
\qquad \XR_{g^i}(\tilde{g}) = \tilde{R}^T_{\tilde{g}} \parcial{g^i}\,,
\ee

\ni where the tilde refers to operations and elements in \Gt. The left- (and
right-, since the U(1) subgroup is central in \Gt) invariant vector field
which at the identity is written as $\parcial{\z}$ is 
$\Xi(\tilde{g})\equiv i\z\parcial{\z}$. It is the vertical (or fundamental) 
vector field associated with the fibre bundle
\be   U(1) \rightarrow \Gtm \rightarrow G \,.
\ee
 
Analogous considerations can be made for the sets of left- and right-invariant
1-forms associated with the set of local canonical coordinates $\{g^i\}$,
i.e. those which at the identity are written as $dg^i$:
\be  
\tilde{\theta}^L{}^{g^i}(\tilde{g})=\tilde{L}^*_{\tilde{g}^{-1}}dg^i = 
 \theta^L{}^{g^i}(g) \,,
\qquad
\tilde{\theta}^R{}^{g^i}(\tilde{g})=\tilde{R}^*_{\tilde{g}^{-1}}dg^i = 
\theta^R{}^{g^i}(g)\,.
\ee

For simplicity of notation, we shall omit the point in which the vector
fields and 1-forms are calculated. Due to the left and right invariance, we 
have $\theta^L{}^{g^i} (\XL_{g^j})=\delta^i_j = \theta^R{}^{g^i}(\XR_{g^j})$.

We can also compute the left- and right-invariant 1-forms which are dual to 
the vertical generator $\Xi$:
\be
\tilde{\theta}^{L\z}=\tilde{L}^*_{\tilde{g}^{-1}} d\z \,,
\qquad 
\tilde{\theta}^{R\z}=\tilde{R}^*_{\tilde{g}^{-1}} d\z \,.
\ee

%

We shall call $\Theta\equiv \tilde{\theta}^L{}^{\z}=\frac{d\z}{i\z} + 
 \frac{\partial \xi(g',g)}{\partial g^i}|_{g'=g^{-1}} dg^i$ the 
{\it quantization 1-form}. It defines a connection on the fibre bundle \Gt 
and it is uniquelly
determined by the two-cocycle $\xi(g_1,g_2)$ (it does not change under changes
of local canonical coordinates of $G$). Adding to $\xi$ a coboundary 
$\xi_{\lambda}$, generated by the function $\lambda$, results in a new 
quantization 1-form $\Theta' = \Theta + \Theta_{\lambda}$ with
\be
\Theta_{\lambda} = \lambda^0_i \theta^L{}^{g^i} - d\lambda \,,
\ee

\ni where $\lambda^0_i\equiv \frac{\partial \lambda(g)}{\partial g^i}|_{g=e}$,
that is, the gradient of $\lambda$ at the identity with respect to the local 
canonical coordinates. Note that $\lambda^0_i$ are constants, and 
therefore, up to the total diferential $d\lambda$, $\Theta_{\lambda}$ is
left invariant under $G$. We also have 
$d\Theta_\lambda = \lambda^0_i d\theta^L{}^{g^i}$, so that using the relation
\be
d\theta^L{}^{g^i} = - \medio C_{jk}^i \theta^L{}^{g^j} \wedge \theta^L{}^{g^k}\,, 
\ee

\ni where $C_{jk}^i,\,i,j,k=1,\ldots, {\rm dim} G$ are the structure 
constants of the Lie algebra ${\cal G}\equiv T_e G$ in the basis of the 
left-invariant
vector fields associated with the set of local canonical coordinates 
$\{g^i\}$, we obtain:
\be
d\Theta_{\lambda} = 
- \medio \lambda^0_i C_{jk}^i \theta^L{}^{g^j}\wedge \theta^L{}^{g^k}\,.
\ee

Note that $\lambda$ defines an element $\lambda^0$ of the coalgebra
${\cal G}^*$ of $G$ characterizing the presymplectic form 
$d\Theta_{\lambda}\equiv d\Theta_{\lambda^0}$. It is easy to see then that
given $\lambda'{}^0$ and $\lambda^0$ on the same orbit of the 
coadjoint action of $G$, $\lambda'{}^0= Ad(g)^*\lambda^0$, for 
some $g\in G$, the corresponding presymplectic forms are related through:
\be
d\Theta_{\lambda'{}^0} \equiv d\Theta_{Ad(g)^*\lambda^0} =
Ad(g)^* d\Theta_{\lambda^0}\,.
\ee

Summarizing this section, we can classify the central extensions of $G$ in 
equivalence classes, using two kinds of equivalence relations. The first one 
is the standard one which leads to the $2^{nd}$ cohomology group 
$H^2(G,U(1))$, where two two-cocycles are cohomologous if they differ by a 
coboundary. According to this, we associate with $\xi$ the class $[[\xi]]$, 
the elements of which differ in a coboundary generated by an arbitrary
function on $G$. 
With this equivalence class
we can associate a series of parameters, given by the corresponding element
of $H^2(G,U(1))$, and will be called the 
{\it cohomology parameters}. An example of them is the mass parameter which
characterizes the central extensions of the Galilei group. However, the 
previous considerations suggest that each equivalence class $[[\xi]]$ should 
be further partitioned according to what we have called pseudo-cohomology 
classes, $[\xi]$, the elements of which differ in a coboundary 
$\xi_{\alpha}$ generated by a function $\alpha$ on $G$ having trivial 
gradient at the identity. Pseudo-cohomology classes are then characterized
by coadjoints orbits of ${\cal G}^*$ which satisfy the integrality condition 
\cite{Kirillov} (the condition of integrality is associated with the 
globality of the generating function $\lambda$ on the group).  


\section{Group Approach to Quantization}
\label{GAQ}
The original Group Approach to Quantization deals with a connected Lie group
which is also a principal $U(1)$-bundle, either a central extension or not. The 
vertical subgroup, $U(1)$, simply realizes the well-established phase invariance
in Quantum Mechanics through the $U(1)$-equivariance condition, on complex 
functions defined on \Gt, intended to restrict this space of functions to the 
linear space of sections of an associated line bundle. The whole 
construction is defined in terms of canonical differential structures on the 
Lie group. However, the quantization formalism can be generalized in two 
directions. On the one hand, the structure group can be replaced by a larger, 
non-necessarily abelian group $T$ containing $U(1)$. In that case, operators 
in $T$, other than the standard $U(1)$, play the role of constraints as we 
generalize, accordingly, the equivariance condition \cite{Ramirez}. On the 
other hand, a non-connected group can substitute \Gt. To proceed in this way 
we must replace all infinitesimal operations on the functions on the group by 
their finite counterparts. This generalization allows us to deal with groups 
with arbitrary homotopy and therefore to quantize systems the configuration 
space (or phase space) of which are not simply connected \cite{Frachall}. We
shall be concerned here with $U(1)$-principal bundles on a connected base. 


\subsection{$U(1)$-quantization}

\ni Let us therefore consider a Lie group \Gt which is a $U(1)$-principal bundle 
with the bundle projection $\pi:\Gtm \rightarrow G$ being a group 
homomorphism. We denote by $\Theta$ the connection 1-form constructed as
explained earlier. It satisfies $i_{\Xi}\Theta=1,\,\,L_{\Xi}\Theta=0$, 
$\Xi$ being the infinitesimal generator
of $U(1)$, or the fundamental vector field of the principal bundle, which
is in the centre of the Lie algebra $\tilde{\cal G}\equiv T_e \Gtm$.
Since $\Theta$ is left-invariant it will be preserved ($L_{\XR}\Theta=0$)
by all right-invariant vector fields (generating finite left 
translations) on $\tilde{G}$. These vector fields are candidates to be
infinitesimal generators of unitary transformations. To determine the 
space of functions on which they should act, we first select complex
valued equivariant functions by requiring
\begin{equation}
L_{\Xi}\psi=i\psi\,. \label{equiv0}
\end{equation}

\ni In other terms, we start with $\Psi: \Gtm \rightarrow C$ and impose the 
equivariance condition (\ref{equiv0}) which identifies them as sections of an associated
line bundle (see, for instance, \cite{Bal}).
To make the action of the right-invariant vector fields on them
irreducible, we have to select appropriate subspaces, and this will be 
achieved by polarization conditions, 

\begin{equation}
L_{\tilde{X}^L}\psi=0\,,\,\,\forall \tilde{X}^L\in {\cal P}\,,
\end{equation}

\ni where the polarization subalgebra ${\cal P}$ is to be discussed at length.
  
The 2-form $\tilde{\Sigma}\equiv d\Theta$ is left invariant under \Gt, and
projectable onto a left-invariant 2-form $\Sigma$ of $G$. 
This, evaluated at the identity,  defines a two-cocycle on the Lie algebra 
$\cal G$.

On vector fields ${\cal X}(G)$ we can define a 
``generalized Lagrange bracket'' \cite{Marmo} by setting, for any pair of 
vector fields $X,Y\in {\cal X}(G)$,
\begin{equation}
(X,Y)_{\Sigma} = \Sigma(X,Y)\in {\cal F}(G)\,. 
\end{equation}

In particular, when we consider left-invariant vector fields 
$X^L,Y^L\in {\cal X}^L(G)$, we get a  real valued bracket:
\begin{equation}
(X^L,Y^L)_{\Sigma} = \Sigma(X^L,Y^L)\in R\,. 
\end{equation}

By evaluating $\tilde{\Sigma}$ at the identity of the group, i.e. on 
$T_e \Gtm=\tilde{\cal G}$,
we can bring it to normal form, which would be the analog of a Darboux frame in
the space of left-invariant 1-forms. We can write
\begin{equation}
\tilde{\Sigma} = \sum_{a=1}^{k}\theta^L{}^a\wedge\theta^L{}^{a+k}\,,
\end{equation}

\ni where $\theta^L{}^a,\theta^L{}^{a+k},\, a=1,\ldots,k$ are left-invariant 1-forms.
We can define a (1,1)-tensor field $J$, a partial (almost) complex structure,
by setting:
\begin{eqnarray}
J\theta^L{}^a &=& \theta^L{}^{a+k} \nn \\
J\theta^L{}^{a+k} &=& -\theta^L{}^a \\
J\theta^L{}^l &=& 0 \nn\, ,
\end{eqnarray}

\ni where $\theta^L{}^l$ are the remaining elements of a basis of left-invariant
1-forms not appearing in $\tilde{\Sigma}$ (that is, those related to 
$Ker\tilde{\Sigma}$.). We also have a ``partial metric tensor'' $\rho$ by
setting $\rho(\theta^L{}^a,\theta^L{}^{a'})=\delta_{aa'},\, 
\rho(\theta^L{}^a,\theta^L{}^l)=0,\,\rho(\theta^L{}^l,\theta^L{}^{l'})=0$.

Usually our considerations will be restricted to finite-dimensional Lie
groups or infinite-dimensional ones possessing a countable basis of generators
for which, for arbitrary fixed $\tilde{X}^L, \, \tilde{\Sigma}(\tilde{X}^L,
\tilde{Y}^L)=0$ except for a finite number of vector fields $\tilde{Y}^L$
(finitely non-zero two-cocycle), and therefore this partial (almost) complex 
structure $J$ can always be introduced.

It is possible to associate with $\Theta$ an horizontal projector, a 
(1,1)-tensor field. We first define the vertical projector 
$V_{\Theta}(X)=\Theta(X) \Xi$, and then $H_{\Theta}=I-V_{\Theta}$.

The characteristic module of $\Theta$ is defined as the intersection 
of $Ker\Theta$ and $Ker d\Theta=Ker\tilde{\Sigma}$. It is generated by a subalgebra of
${\cal X}^L(\Gtm)$, the characteristic subalgebra $\GC$. Elements in 
$Ker\Theta\cap Ker d\Theta$ are easily shown to be a Lie algebra. In fact, 
it follows from the identity 
$d\Theta(X,Y)=L_X\Theta(Y)-L_Y\Theta(X) -\Theta([X,Y])$.

It turns out that $P\equiv \Gtm/{\cal G}_C$ is a quantum 
manifold in the sense of Geometric Quantization, with connection the projection
of $\Theta$ to $P$ (see \cite{23}), and $d\Theta$ projected onto 
$P/U(1)$ is a symplectic 2-form. This establishes the connection with the 
Coadjoint Orbits Method, the different coadjoint orbits being obtained by
suitable choice of the (pseudo-)extension parameters. 

A {\bf first-order polarization} or just {\bf polarization} ${\cal P}$ is 
defined as a maximal horizontal 
left subalgebra. The horizontality condition means that the polarization 
is in $Ker\Theta$. Again, by using the identity 
$d\Theta(X,Y)=L_X\Theta(Y)-L_Y\Theta(X) -\Theta([X,Y])$ we find that the
generalized Lagrange braket  of any two elements of ${\cal P}$ vanishes. 
Therefore we find that a polarization is an isotropic maximal subalgebra.
We notice that maximality refers to the Lie commutator (subalgebra)
and not to isotropy (Lagrange bracket). 

A polarization may have non-trivial intersection with the characteristic 
subalgebra. 
We say that a polarization is {\bf full (or regular)} if it contains the 
whole characteristic subalgebra. 
We also say that a polarization $\cal P$ is 
{\bf symplectic} if $\tilde{\Sigma}$ on ${\cal P} \oplus J{\cal P}$ is 
of maximal rank [When the quantization 1-form $\Theta$ is associated with a 
co-adjoint orbit of $G$, full polarizations satisfy Pukanszky's condition, 
and full and symplectic polarizations correspond to ``admissible" 
subalgebras subordinated to $\Theta|_e$ \cite{Kirillov}].

It should be stressed that the notion of polarization and characteristic 
subalgebras here given in terms of $\Theta$ is really a consequence 
of the fibre bundle structure of the group law of \Gt and, therefore, 
can be translated into finite (versus infinitesimal) form defining 
the corresponding subgroups (see \cite{Ramirez}).
  
{}From the geometric point of view, a polarization defines 
a foliation via the Frobenius theorem. It is possible to select subspaces of 
equivariant complex-valued functions on \Gt, by requiring them to be constant 
along integral leaves of the foliation associated with the polarization.
Whether this subspace is going to carry an irreducible representation
for the right-invariant vector fields is to be checked. When the polarization
is full and symplectic we get leaves which are maximally isotropic 
submanifolds for $d\Theta$. The selected subspaces of equivariant complex-valued 
functions on \Gt, which we may call wave functions, will be 
characterized by $L_{\Xi}\Psi=i\Psi,\, L_{X^L}\Psi=0,\,\forall X^L\in 
{\cal P}$.

Finally, the Hilbert space structure on the space of wave functions is
provided by the invariant Haar measure constituted by the exterior product of 
the components of the left-invariant canonical 1-form: $\mu=
\theta^{Lg^1}\wedge\theta^{Lg^2}\wedge...$. The finiteness of the scalar 
product will eventually restrict the values of the (pseudo-)extension 
parameters that characterize the representations. 

The classical theory for the system is easily recovered by defining the
Noether invariants as $F_{g^i}\equiv i_{\XR_{g^i}} \Theta$. A Poisson 
bracket can be introduced (defined by $d\Theta$), in such a way that the
Noether invariants generate a Lie algebra isomorphic (if there are no 
algebraic anomalies, see Sec. 3) to that of \Gt (see \cite{23} for a complete 
description of the classical theory using the GAQ formalism).


\subsection{Non-horizontal polarizations}

\ni We can generalize the notion of polarization introduced above by relaxing 
the condition of horizontality, and define a {\bf non-horizontal 
polarization} as a maximal left subalgebra of $\tilde{\cal G}^L$ not containing 
the vertical generator $\Xi$. Although this seems to be a more general notion, 
the following proposition states that it is related to the previous one.

{\bf Proposition}: Given a Quantization 1-form $\Theta$ and a non-horizontal
polarization ${\cal P}$ with respect to it, it is always possible to find
a left-invariant 1-form $\Theta'=\Theta + \alpha_i\theta^L{}^{g^i}$ for which 
$\cal P$ is horizontal.

{\it Proof}: Let $\{X^L_{(k)}\}_{k=1}^m$ be a basis for $\cal P$, where 
$m={\rm dim} {\cal P}$, and denote $a^0_{(k)} \equiv \Theta(X^L_{(k)})$, which
are, in general, non-zero for strictly non-horizontal polarizations. The
condition for $\cal P$ of being horizontal with respect to $\Theta'$ is written 
as:
\be 
\sum_{i=1}^{n} \alpha_i a^i_{(k)} = - a^0_{(k)}\,,\qquad k=1,\ldots,m \,,
\ee

\ni where $a^i_{(k)} \equiv \theta^L{}^{g^i}(X^L_{(k)})$, and $n={\rm dim} {\cal G}$.
This is an undetermined equation for the $\alpha_i\,,i=1,\ldots,n$, and the
dimension of the space of solutions, since $X^L_{(k)}$ form a basis for
$\cal P$, is ${\rm dim} {\cal G} - {\rm dim} {\cal P}$.

Note that, up to an (irrelevant) total differential, $\Theta'$ coincides
with the quantization 1-form obtained after adding a coboundary generated by
a function $\lambda$ with $\lambda^0_i = \alpha_i$. Therefore, polarizing with a
non-horizontal polarization is equivalent to introducing a pseudo-cocycle in the
group law (and polarize with respect to a horizontal polarization), and, 
accordingly, non-horizontal polarizations can be classified according to 
pseudo-cohomology classes. This equivalence will be exploited in some of the
examples. In addition, in generalizing first-order polarizations
to higher-order ones, the notion of non-horizontal polarization will turn out to be a more
appropiate one to start with.

\section{Simple Examples}

\subsection{The abelian group $R^{k}$}
\label{H-W}

The simplest example one can think of is the abelian group 
$R^k$, with $k>1$ (the case $k=1$ is trivial, since it admits no non-trivial
symplectic structure). Since the coadjoint action is trivial, all its 
coadjoint orbits are zero dimensional. This means that there will be no 
pseudo-cohomology classes, and only true central extensions have to be
considered.

Given any two-cocycle $\xi$ defining a central extension $\tilde{R}^k$ of 
$R^k$, then $d\Theta$, where $\Theta$ is the quantization 1-form, is always 
left- (and right- in this case) invariant and exact, but it is not invariantly
exact (as a 2-form on $R^k$), 
due to the non-trivial group cohomology of $R^k$. If we parametrize $R^k$ 
with (global) canonical coordinates $\x = (x_1,x_2,...,x_k)$, then, since it
is left invariant and the group is abelian, $d\Theta$ can be written as:
\be  d\Theta= a_{ij} dx^i\wedge dx^j\,,
\ee
 
\ni where $a_{ij}$ is an antisymmetric $k\times k$ constant matrix. We can 
choose a
two-cocycle representative of the associated cohomology class of the form: 
\be \xi(\x_1,\x_2) = a_{ij} x^i_1 x^j_2\,,
\ee

\ni any other representative will differ from this by a coboundary that,
due to the trivial pseudo-cohomology of $R^k$, will contribute to $\Theta$
with an irrelevant total differential.
Therefore, $\xi$ is an anti-symmetric bilinear function on $R^k$, and with
an appropriate change of coordinates in $R^k$ can be taken to normal form, 
in which the matrix $a_{ij}$ is written as:
\be
\medio \left( \begin{tabular}{c|c|c}
$0_n$& $D_n$ & \\
\rule{2cm}{0.15mm}&  \rule{2cm}{0.15mm}&  $0_{2n\times r}$ \\
$-D_n$ & $0_n$ & \\
\rule{2cm}{0.15mm}&  \rule{2cm}{0.15mm} &  \rule{2cm}{0.15mm} \\
\multicolumn{2}{c|}{$0_{r\times 2n}$} &  $0_r$ 
\end{tabular} \right)\,,
\ee

\ni where $0_p$ is the $p\times p$ zero matrix, $0_{p\times q}$ is the zero 
$p\times q$ matrix, and $D_n$ is a $n\times n$ real matrix of 
the form:
\be \left( \begin{array}{cccccc}
 \nu_1 & 0 & . & .& .& 0 \\
0 & \nu_2& . & . & .& 0 \\
. & . & .& .& .& . \\
. & . & .& .& .& . \\
. & . & .& .& \nu_{n-1} & 0 \\
0 & . & .& .& 0& \nu_n 
\end{array} \right)\,,
\ee

\ni with $k=2n+r$. The parameters $\nu_1,\ldots, \nu_n$ characterize the
extension $\tilde{R}^k$, and, thus, they are the cohomology parameters. 
In physical situations, the subspace $R^{2n}$ of $R^k$ is the tangent  
space $T(R^n)$ of a physical system (at this point there is no distinction 
between the tangent or the phase space $T^*(R^n)$, and we shall consider it the 
tangent space for convenience).  
The requirement of isotropy (under spatial rotations) will fix these 
parameters to coincide, $\nu_i=\nu,\forall i=1,\ldots,n$. 
In this case, the two-cocycle can be written as:
\be
\xi(\q_1,\vec{v}_1,\a_1;\q_2,\vec{v}_2,\a_2) = 
\medio \nu (\q_2\cdot\vec{v}_1-\q_1\cdot\vec{v}_2)\,,
\ee

\ni where $\q_{1,2}$ are $n$-dimensional vectors corresponding to the first 
$n$ coordinates (in the new basis), $\vec{v}_{1,2}$ correspond to the following $n$ 
``conjugated" coordinates,
and $\a_{1,2}$ to the remaining $r$ coordinates. 
Since the two-cocycle $\xi$ does not depend on
the coordinates $\a_{1,2}$, the group $\tilde{R}^k$ can be written as 
H-W$_n\times R^r$, where H-W$_n$ is the well-known Heisenberg-Weyl group with $n$ coordinates
and $n$ velocities. 
The group law for $\tilde{R}^k$ can be rewritten in terms of the new 
coordinates:
\begin{eqnarray}
\q\,'' & = & \q\,' + \q \nn \\
\v\,'' & = & \v\,' + \v  \\
\a\,'' & = & \a\,' + \a \nn \\
\z'' & = & \z'\z e^{\frac{i}{2} \nu (\q\,'\v - \v\,'\q)} \nn \,.
\end{eqnarray} 

{}From the group law we see that if $\q$ is interpreted as coordinates, and 
$\vec{v}$ as velocities, then $\nu=\frac{m}{\hbar}$. Therefore, the cohomology 
parameter for the (isotropic) Heisenberg-Weyl group can be identified with
$\frac{m}{\hbar}$. The variables $\a$ do not play any role, and can be 
factorized, as we shall see later.

Left- and right-invariant (under the group $\tilde{R}^k$) vector fields are:
\begin{equation}
\begin{array}{rcl}
\XL_{\q} & = & \parcial{\q} - \frac{m}{2\hbar} \vec{v} \Xi \\
\XL_{\v} & = & \parcial{\v} + \frac{m}{2\hbar} \q \Xi \\
\XL_{\a} & = & \parcial{\a}  \\
\end{array}\qquad
\begin{array}{rcl}
\XR_{\q} & = & \parcial{\q} + \frac{m}{2\hbar} \v \Xi \\
\XR_{\v} & = & \parcial{\v} - \frac{m}{2\hbar} \q \Xi \\
\XR_{\a} & = & \parcial{\a} \,, \\
\end{array}\,
\end{equation}

\ni and $\Xi=i\z\parcial{\z}$ is the vertical (left- and right-invariant) 
vector field. The commutation relations for these vector fields are:
\be
[\XL_{q^i},\XL_{v^j}] = \frac{m}{\hbar} \Xi\,,
\ee

\ni the rest of them being zero. Left and right invariant 1-forms for $R^k$ are 
simply $d\x$, for $\x=\q,\v$ and $\a$. The quantization 1-form $\Theta$, 
which for convenience we redefine with a factor $\hbar$, is:
\be
\Theta = \hbar\frac{d\z}{i\z} + \frac{1}{2}(m\v\cdot d\q - \q\cdot d(m\v))\,. 
\ee

Note that $d\Theta =  d(m\v)\wedge d\q$ is a pre-symplectic form on
$R^k$, with kernel the subspace $R^r$ spanned by the vectors $\a$. On the
quotient $R^k/R^r = R^{2n}$, $d\Theta$ is a true symplectic form.
In fact, a partial complex structure $J$ can be introduced, of the form
$J= d(mv^i) \otimes \XL_{q^i} - \frac{1}{m} dq^i\otimes \XL_{v^i}$ (that is, $J$ satisfies
$J(\XL_{q^i})= -\frac{1}{m} \XL_{v^i}$ and $J(\XL_{vi})= m\XL_{q^i}$). $J$
turns out to be a complex structure on the reduced space $R^k/R^r$.
The characteristic subalgebra, i.e. $Ker \Theta\cap\ Ker d\Theta$, 
is accordingly given by ${\cal G}_{\Theta} = <\XL_{\a}>$.

The possible horizontal polarizations for this group are of the form:
\be
{\cal P} = < \XL_{\a}, \alpha_i^j \XL_{q^j} + \beta_i^j\XL_{v_j},
            i=1,\ldots,n> \,, 
\ee

\ni with restrictions on the real coefficients $\alpha_i^j,\beta_i^j$ making 
it maximal, and horizontal with respect to $\Theta$. These 
restrictions are necessary for the polarizations to be full and symplectic.
If, again, one imposes isotropy under rotations (if rotations are taken into 
account they must be included in the polarization), the coefficients 
$\alpha_i^j$ and $\beta_i^j$ must be proportional.

 There are two
``natural" polarizations, ${\cal P}_p = <\XL_{\a}, \XL_{\q}>$ and
${\cal P}_q = <\XL_{\a}, \XL_{\v}>$, leading to the representations in 
momentum and configuration space, respectively. It should be stressed that 
these polarizations lead to equivalent representations of $R^k$, and 
the unitary operator relating the representations obtained 
with  ${\cal P}_p$ and ${\cal P}_q$ is the Fourier transform. This is an 
outer isomorphism of H-W$_n$, but it is inner in the Weyl-Symplectic group 
$WSp(2n,R)$ \cite{Wolf}, as we shall see later.

Taking advantage of the natural complex structure of $R^{2n}\approx C^n$ 
(the one induced by $J$), we can choose a complex polarization of the form:
\be
{\cal P}_c = <\XL_{\a}, \XL_{\q} + i\w\XL_{\v}>\,,
\ee

 \ni where $\w$ is a constant with the appropriate dimensions (from the
physical point of view, it will be frequency, which makes
this polarization appropriate for the description of the Harmonic 
Oscillator; see Sec. \ref{Chorri}). This polarization leads to a 
representation in terms of 
holomorphic (or anti-holomorphic) functions on $C^n$. It is unitarily 
equivalent to the other representations, the unitary transformation which
relates it with the representation in configuration space being the Bargmann
transform. This is also an outer automorphism of H-W$_n$, but it is inner in 
a certain subsemigroup of $Sp(2n,C)$ \cite{Wolf}.

Let us compute the representation obtained with the 
polarization ${\cal P}_q$. The equations $\XL_{\a} \Psi = 0$ leads to
wave functions not depending on the $\a$ variables (they trivially 
factorize and we can forget about them), and the equations $\XL_{\v}\Psi =0$ 
(together with the equivariance condition $\Xi\Psi = i\Psi$) lead to:
\be
\Psi = \z e^{-\frac{im}{2\hbar} \q\cdot\v} \Phi(\q\,)\,,
\ee

\ni where $\Phi(\q)$ is an arbitrary function of $\q$ (appart from 
normalizability considerations). If we compute the action of the 
right-invariant vector fields on these wave functions, we obtain:
\begin{eqnarray}
\XR_{\q} \Psi &=& \z e^{-\frac{im}{2\hbar} \q\cdot\v} 
               \parcial{\q}\Phi(\q) \nn \\
\XR_{\v} \Psi &=& -\z e^{-\frac{im}{2\hbar} \q\cdot\v} 
               \frac{im}{\hbar}\q\Phi(\q)\,.
\end{eqnarray}

\ni This representation is unitarily equivalent to the Schr\"odinger 
representation (labeled by the values of the cohomology parameter 
$\frac{m}{\hbar}$) of the H-W group, and the unitary operator relating them is the
operator $e^{-\frac{im}{2\hbar} \q\cdot\v}$. As expected, 
it reproduces the standard Weyl commutation relations, the basis of 
non-relativistic Quantum Mechanics.

With the polarization ${\cal P}^{HO}_v$ we obtain the representation in
velocity (momentum) space, with wave functions of the form:
\begin{equation}
\Psi = \z e^{\frac{im}{2\hbar}\q\cdot\v}\Phi(\v)\,,
\end{equation}

\ni where $\Phi(\v)$ is an arbitrary function of $\v$. The action of 
the right-invariant vector fields are:
\begin{eqnarray}
\XR_{\q} \Psi &=& \z e^{\frac{im}{2\hbar} \q\cdot\v} 
              \left(\frac{im}{\hbar}\v\right)\Phi(\v) \nn \\
\XR_{\v} \Psi &=& \z e^{\frac{im}{2\hbar} \q\cdot\v} 
                 \left(\parcial{\v}\right)\Phi(\v) \,.               
\end{eqnarray}

If we now solve for the holomorphic polarization ${\cal P}_c$, we obtain, 
introducing the appropriate (complex) change of variables 
$\c \equiv \frac{1}{\sqrt{2}}(\q + i\v/\w)$ and its complex conjugate, the
following wave functions:
\be
\Psi = e^{\frac{m\w}{2\hbar} \c\cdot\c^*} \Phi(\c)\,,
\ee

\ni where $\Phi(\c)$ is an arbitrary holomorphic function. The action of the 
right-invariant vector fields is:
\begin{eqnarray}
\XR_{\c}\Psi &=& e^{\frac{m\w}{2\hbar} \c\cdot\c^*} 
               \left(\frac{m\w}{\hbar} \c\right) \Phi(\c) \nn \\
\XR_{\c^*}\Psi &=&  e^{\frac{m\w}{2\hbar} \c\cdot\c^*} 
                \left(\parcial{\c} \right)\Phi(\c) \,, 
\end{eqnarray}
 
\ni leading to the standard Bargmann's holomorphic representation. The 
generators $\XR_{\c}$ and its complex conjugate have the usual form 
$\frac{1}{\sqrt{2}}\left( \XR_{\q} \mp\w \XR_{\v}\right)$.

Finally, we shall consider another real polarization, of the form
$< \XL_{\a}, \XL_{\q} + \w \XL_{\v}>$. Introducing the adequate change of 
variables, $\r_{\pm} \equiv \frac{1}{\sqrt{2}} (\q \pm \v/\w)$, the 
polarization equations lead to the wave functions:
\be 
\Psi = e^{-i\frac{m\w}{2\hbar} \r_{+}\cdot\r_{-}} \Phi(\r_{-})\,,
\ee 

\ni where $\Phi(\r_{-})$ is an arbitrary function of $\r_{-}$. As we shall 
see later, this representation is adequate for the description of the 
{\it repulsive Harmonic Oscillator}, see Sec. \ref{Chorri}. The action 
of the right-invariant vector fields are:
\begin{eqnarray}
\XR_{\r_{+}} \Psi &=& -e^{-i\frac{m\w}{2\hbar} \r_{+}\cdot\r_{-}} 
         \left(\frac{im\w}{\hbar} \r_{-}\right) \Phi(\r_{-}) \nn \\
\XR_{\r_{-}} \Psi &=& e^{-i\frac{m\w}{2\hbar} \r_{+}\cdot\r_{-}}
            \left(\parcial{\r_{-}}\right) \Phi(\r_{-})\,, 
\end{eqnarray}

\ni where $\XR_{\r_{\pm}} \equiv \frac{1}{\sqrt{2}} 
             \left( \XR_{\q} \pm \w \XR_{v}\right)$.

 In this simple example no temporal evolution is considered. This question
will be addressed in Sec. \ref{Chorri}, with the example of the Schr\"odinger group, 
the group of linear canonical transformation acting on the H-W group. 
Different choices of ``time" in it (among different uniparametric subgroups of
$SL(2,R)$) will provide us with different dynamical systems, like the
free particle, the harmonic oscillator and the repulsive oscillator 
(see \cite{Wolf}). Each time generator will select, among the ones considered here, the 
appropriate (invariant) polarization.


\subsection{The semisimple group $SU(2)$}
\label{SU(2)}

Let us consider now an example, which in a certain sense is on the other 
extreme to that
of the abelian group $R^k$. It is the semisimple group $SU(2)$, which has
trivial cohomology group, $H^2(SU(2),U(1))=0$. For this reason all 
two-cocycles on $SU(2)$ are coboundaries, and they will be classified 
according to pseudo-cohomology classes only.

A group law for $SU(2)$ can be obtained making use of its realization as 
$2\times 2$ complex matrices of the form:
\be
\left(\begin{array}{cc}
z_1 & z_2 \\
-z_2^* & z_1^* 
\end{array} \right)\,,
\ee

\ni with $|z_1|^2 + |z_2|^2 =1$, with matrix multiplication as group law.

We shall proceed in an indirect way in order to keep global coordinates. Let us relax the 
condition $|z_1|^2 + |z_2|^2 =1$ to $|z_1|^2 + |z_2|^2 > 0$, and apply the formalism to 
this group.

The group law, obtained from the multiplication of matrices, is:
\begin{eqnarray}
z_1'' & = & z_1'z_1 -z_2'z_2^* \nn \\
z_2'' & = & z_1'z_2 + z_2'z_1^* \\
z_1^*{}'' & = & (z_1'')^* \nn \\
z_2^*{}'' & = & (z_2'')^* \nn \,.
\end{eqnarray}

Left- and right-invariant vector fields are particularly simple:
\be 
\begin{array}{lcl}
X^L_{z_1} & = & z_1\parcial{z_1} + z_2^*\parcial{z_2^*} \\
X^L_{z_2} & = & z_1\parcial{z_2} + z_2^*\parcial{z_1^*} \\
X^L_{z_1^*} & = & z_1^*\parcial{z_1^*} + z_2\parcial{z_2} \\
X^L_{z_2^*} & = & z_1^*\parcial{z_2^*} + z_2\parcial{z_1} 
\end{array}
\qquad\qquad
\begin{array}{lcl}
X^R_{z_1} & = & z_1\parcial{z_1} + z_2\parcial{z_2} \\
X^R_{z_1} & = & z_1^*\parcial{z_2} - z_2^*\parcial{z_1} \\
X^R_{z_1^*} & = & z_1^*\parcial{z_1^*} + z_2^*\parcial{z_2^*} \\
X^R_{z_2^*} & = & z_1\parcial{z_2^*} - z_2\parcial{z_1^*}\,,
\end{array}
\ee
 
\ni and the commutation relations for left-invariant vector fields are:
\be
\begin{array}{lcl}
\left[X^L_{z_1},X^L_{z_2}\right] & = & X^L_{z_2} \\
\left[X^L_{z_1},X^L_{z_1^*}\right] & = & 0 \\
\left[X^L_{z_1},X^L_{z_2^*}\right] & = & -X^L_{z_2^*} 
\end{array}
\qquad\qquad
\begin{array}{lcl}
\left[X^L_{z_1^*},X^L_{z_2}\right] & = & - X^L_{z_2} \\
\left[X^L_{z_1^*},X^L_{z_2^*}\right] & = & X^L_{z_2^*}\\
\left[X^L_{z_2},X^L_{z_2^*}\right] & = & - X^L_{z_1} + X^L_{z_1^*}\,, 
\end{array}
\ee

Note that the combination $X^L_{z_1} + X^L_{z_1^*}$ is a central generator, 
commuting with all the other vector fields. It is a gauge \cite{gauge} 
generator, and it will be contained in any polarization. Therefore its action
(since it is central, it is left- and right-invariant) on the wave functions
is always trivial. Thus, our Lie algebra decomposes in the direct sum 
$su(2) \oplus R$, where $su(2)$ is generated by $X^L_{z_1} - X^L_{z_1^*}, 
X^L_{z_2}$ and $X^L_{z_2^*}$.
In this way we return to our original problem of quantizing $SU(2)$,
but with the advantage of using the global coordinates $z_1$ and $z_2$, which 
will allow us to use finite transformations (not reachable with local 
coordinates ) to implement the compatibility of charts.

 Now we should consider central extensions of $SU(2)$, which, due to its 
trivial group cohomology, will be pseudo-extesions, classified according to
pseudo-cohomology classes. However, in order to illustrate the equivalence 
between pseudo-extensions and non-horizontal polarizations, we shall consider
the direct product $SU(2)\times U(1)$ (trivial pseudo-extension) and introduce
an appropriate non-horizontal polarization. 

In this case the group law for the variable $\z\in U(1)$ is simply 
$\z''=\z'\z$ and the generator associated with it, which is both left- and
right-invariant, is $\Xi = i\z\parcial{\z}$. The quantization 1-form is 
$\Theta= \frac{d\z}{i\z}$. Since the extension is trivial, the whole group $SU(2)$ 
(times $R$, generated by the gauge generator $X^L_{z_1}+X^L_{z_1^*}$) has its infinitesimal 
generators in 
the caracteristic subalgebra. A horizontal polarization would be constituted 
by the caracteristic subalbegra, and it would lead, obviously, to the trivial
representation of $SU(2)$, with spin zero. 

Let us introduce, instead, a non horizontal polarization of the form:
\be
 {\cal P}^{n.h.} = < X^L_{z_1}+X^L_{z_1^*},X^L_{z_1}-X^L_{z_1^*} - 
\lambda \Xi,
 X^L_{z_2} > \,,
\ee

\ni where $\lambda \in R$ is a parameter characterizing the polarization. 
Later, we shall see that the integrability of the polarization 
equations to the whole group will 
restrict the possibles values of $\lambda$.
Solving the first polarization equation, $(X^L_{z_1}+X^L_{z_1^*})\Psi =0$,
which states the gauge caracter of this generator, leads to a wave function
defined on $SU(2)$,
\be 
\Psi = \Psi(w_1,w_2,w_1^*,w_2^*)\,,
\ee

\ni where $w_i = \frac{z_i}{\sqrt{ |z_1|^2+|z_2|^2}},\,i=1,2$, in such a way
that $|w_1|^2+|w_2|^2 =1$, as expected. 

The second polarization equation (together with the $U(1)$-equivariance 
condition $\Xi\Psi =i\Psi$), written in terms of the new variables, reads:
\be
\left[ w_1\parcial{w_1} + w_2^*\parcial{w_2^*} - w_1^*\parcial{w_1^*}
- w_2\parcial{w_2} - \lambda \right] \Psi = 0\,,
\ee

\ni while the last polarization condition is written as:
\be
\left[ w_1\parcial{w_2} - w_2^* \parcial{w_1^*} \right]\Psi =0\,.
\ee

 At this point it is convenient to introduce local charts in $SU(2)$ in order 
to solve these polarization equations. Thus, two cases are considered:
\begin{itemize}
\item $z_1\neq 0$: The solutions are of the form: 
\be
\Psi = w_1^{\lambda} \Phi(\tau)\,,
\ee

\ni with $\Phi$ is an arbitrary holomorphic function of the variable 
$\tau \equiv \frac{w_2^*}{w_1}$.
\item $z_2^*\neq 0$: The solutions are of the form:
\be
\tilde{\Psi} = w_2^*{}^{\lambda} \tilde{\Phi}(\tilde{\tau})\,,
\ee

\ni where $\tilde{\Phi}$ is an arbitrary holomorphic function of the variable
$\tilde{\tau} \equiv \frac{w_1}{w_2^*} = \frac{1}{\tau}$ 
\end{itemize}

These solutions are the same that those one obtains considering stereographic projection 
coordinates on the sphere $S^2 \approx SU(2)/U(1)$, but with the advantage that one can pass 
from one chart to the other with the action of the
element $J=\left( \begin{array}{cc} 0&1 \\ -1 &0\end{array} \right)\in SU(2)$, 
which satisfies $J^4=I_2$. However, the repeated action 
of this element on the wave functions is 
$\Psi(J^4*g) = (-1)^{-2\lambda} \Psi(g)$, and, therefore, the requirement of 
single-valuedness of the representation 
(since the group $SU(2)$ is simply connected all its representations are 
single-valued) implies that $\lambda \in Z$. This condition is equivalent 
to the condition of chart compatibility.

 Now we have to compute the action of the right-invariant vector fields on
polarized wave functions. Since we can pass from one chart to the other 
with the aid of the element $J$, we only have to compute it on the chart containing the 
identity of the group:
\begin{eqnarray}
(\XR_{z_1} - \XR_{z_1^*})\Psi &=& w_1^{\lambda} \left[ \lambda \Phi 
- 2\tau \frac{\partial \Phi}{\partial \tau} \right] \nn \\
\XR_{z_2} \Psi &=& w_1^{\lambda} \left[-\lambda \tau + 
 \tau^2 \frac{\partial \Phi}{\partial \tau}\right] \\
\XR_{z_2^*} \Psi &=& w_1^{\lambda} \left[ 
  \frac{\partial \Phi}{\partial \tau}\right] \nn\,.
\end{eqnarray}

It is straightforward to check that $w_1^{\lambda}$ and $w_1^{\lambda}\tau^{\lambda}$ are 
maximal and minimal weight states, respectively. Therefore 
the Hilbert space (on which the right-invariant vector fields act irreducibily) has dimensinon 
$\lambda+1$ and is generated by the wave functions 
$\{ w_1^{\lambda},w_1^{\lambda}\tau, \ldots, w_1^{\lambda}\tau^{\lambda} \}$. Clearly, we can 
identify $\lambda$ with $2j$, $j$ being the spin, which characterizes the irreducible 
representations of $SU(2)$. 


\section{Algebraic Anomalies}

In Sec. 3, we introduced  the concept of full and symplectic 
polarization subalgebra intended to reduce the representation obtained 
through the right-invariant vector fields acting on equivariant functions on
the group. It contains ``half'' of the symplectic vector fields as well as 
the entire characteristic subalgebra. If the full reduction is achieved, 
the whole set of physical 
operators can be rewritten in terms of the basic ones, i.e. those 
which are the right version of the left-invariant generators in 
$J{\cal P}\oplus J^2{\cal P}$. For instance, the 
energy operator for the free particle can be written as 
$\frac{\hat{p}^2}{2m}$, the angular momentum in 3+1 dimensions is the 
vector product ${\bf \hat{x}}\times {\bf \hat{p}}$, or the energy for 
the harmonic oscillator is $\hat{c}^{\dag}\hat{c}$ (note that, since we 
are using first-order polarizations, all this operators are really written
as first-order differential operators).

However, the existence of a full and symplectic polarization is guaranteed
only for semisimple and solvable groups \cite{Kirillov}.
We define  an {\bf anomalous} group \cite{Anomalias} as  a central extension 
\Gt which does not admit any polarization which is full and symplectic 
for some values of the (pseudo-)cohomology parameters, called the 
{\bf classical} values of the anomaly (they are called classical because they
are associated with the coadjoints orbits of the group \Gt, that is, with 
the classical phase space of the physical system).
Anomalous groups feature another set of values of the (pseudo-)cohomology 
parameters, called the {\bf quantum} values of the anomaly, for which the 
carrier space associated
with a full and symplectic polarization contains an invariant subspace.
For the classical values of the anomaly, the classical solution manifold 
undergoes a reduction in dimension thus increasing the number of 
(non-linear) relationships among Noether invariants,  whereas for the quantum 
values the number of 
basic operators decreases on the invariant (reduced) subspace due to
the appearance of (higher-order) relations among the quantum operators.

We must remark that the anomalies we are dealing with in this paper are
of {\it algebraic} character in the sense that they appear at the Lie algebra 
level, and must be distinguished from the {\it topologic anomalies} which are
associated with the non-trivial homotopy of the (reduced) phase space 
\cite{Frachall}.

The non-existence of a full and/or symplectic polarization is traced back to 
the presence in the characteristic subalgebra, for certain values 
of the (pseudo-)cohomology parameters (the classical values of the anomaly), 
of some 
elements the adjoint action of which are not diagonalizable in the 
``$x-p$-like" algebra subspace. The anomaly problem here presented parallels 
that of the
non-existence of invariant polarizations in the Kirillov-Kostant co-adjoint
orbits method \cite{Gotay}, and the conventional anomaly problem in
Quantum Field Theory which manifests itself through the appearance of
central charges in the quantum current algebra, absent from the classical
(Poisson bracket) algebra \cite{Jackiw}.
 
The full reduction of representations in anomalous cases will be achieved 
by means of a generalized concept of (higher-order) polarization (see below). 

Let us try to clarify the situation in terms of coordinates and employing
a rather symbolic language. We may imagine our group $\tilde{G}$ parametrized
by $(\kappa,\pi;\epsilon^+,\epsilon^-,\epsilon^0;\zeta)$ and with Lie algebra 
of the form:
\begin{eqnarray}
\left[\XL_{\kappa},\XL_{\pi}\right]&=&\XL_{\epsilon^0}+a\Xi\;,\;\;a\in R \nn \\
\left[\XL_{\epsilon^+},\XL_{\epsilon^-}\right]&=&\XL_{\epsilon^0}  \label{Anomalia} \\
\left[\XL_{\epsilon^{\pm}},\XL_{\kappa}\right]&=&\alpha^{\pm}\XL_{\kappa}+
 \beta^{\pm}\XL_{\pi} \nn \\
\left[\XL_{\epsilon^{\pm}},\XL_{\pi}\right]&=&\gamma^{\pm}\XL_{\kappa}+
\delta^{\pm}\XL_{pi} 
\nn \,,
\end{eqnarray}

\ni where it is assumed that the adjoint action of none of 
$\XL_{\epsilon^{\pm}}$ is 
diagonalizable (only the action of $\XL_{\epsilon^0}$, here omitted, can be
diagonalized). A glance at (\ref{Anomalia}) reveals the structure of 
$\GC$:
\begin{equation}
\GC=<\XL_{\epsilon^0},\XL_{\epsilon^+},\XL_{\epsilon^-}>\,,
\end{equation} 

\ni and the fact that {\it there is no full and symplectic polarization}; 
only non-full (though symplectic) polarizations ${\cal P}^{\pm}$ are found 
which exclude $\XL_{\epsilon^{\mp}}\in \GC$, and the non-symplectic 
(though full) polarization $\GC$ itself. Quantizing with ${\cal P}^{\pm}$ 
means that $\XR_{\epsilon^{-}}$ and $\XR_{\epsilon^{+}}$ {\it cannot
be expressed in terms of $\XR_{\kappa},\XR_{\pi}$}. The situation is as if 
$\XR_{\epsilon^{\mp}}$ also were basic operators, i.e. as if 
\begin{equation}
\left[\XL_{\epsilon^+},\XL_{\epsilon^-}\right]=\XL_{\epsilon^0}+k\Xi 
\label{commutatork} \,,
\end{equation}

\ni for some numerical constant $k$. Quantizing with the polarization 
${\cal P}_C\equiv\GC$ leads to a rather unconventional representation 
in terms of the $\kappa$'s and $\pi$'s variables, like Van Hove's 
prequantization of quadratic polynomials on phase-space, which is 
reducible, decomposing in two invariant subspaces \cite{Hove}. 

The reasoning above suggests another way of looking at anomalies. Let us start from
a doubly extended group $\tilde{G}$ with initially independent 
(pseudo-) extension parameters $a,k$. This involves adopting the following 
Lie algebra:
\begin{eqnarray}
\left[\XL_{\kappa},\XL_{\pi}\right]&=&\XL_{\epsilon^0}+a\Xi\;,\;\;a\in R \nn \\
\left[\XL_{\epsilon^+},\XL_{\epsilon^-}\right]&=&\XL_{\epsilon^0}+k\Xi\;,\;\;k\in R  \\
\left[\XL_{\epsilon^{\pm}},\XL_{\kappa}\right]&=&\alpha^{\pm}\XL_{\kappa}+
    \beta^{\pm}\XL_{\pi} \nn \\
\left[\XL_{\epsilon^{\pm}},\XL_{\pi}\right]&=&\gamma^{\pm}\XL_{\kappa}+
    \delta^{\pm}\XL_{\pi} \nn \,,
\end{eqnarray}
  
\ni both for classical and quantum dynamics. Accordingly, we should have to
admit that the polarizations mentioned above excluding $\XL_{\epsilon^{\mp}}$:
\begin{eqnarray}
{\cal P}^+&=&<\XL_{c^+},\XL_{\epsilon^0},\XL_{\epsilon^+}> \nn \\
&{\rm or}&   \\
{\cal P}^-&=&<\XL_{c^-},\XL_{\epsilon^0},\XL_{\epsilon^-}> \nn \,,
\end{eqnarray}

\ni where $c^{\pm}$ are linear combinations of $\kappa,\pi$ 
diagonalizing $\XL_{\epsilon^0}$,
are now full and symplectic polarizations (for $k\neq 0$), so that only 
$\XL_{\epsilon^0}$ should be expected to be
expressible in terms of basic operators. However, and this is the remembrance
of the underlying anomaly, one finds that for some specific values of 
$a,k$,
actually arbitrary $a$ and a certain $k=k(a)$, the ``basic" operators 
$\XR_{\epsilon^{\pm}}$
turn out to be rewritten, accidentally, as functions of the originally basic
operators $\XR_{\kappa,\pi}$ or $\XR_{c^{\pm}}$. Those specific values of 
$a,k$ are the {\it quantum values of the anomaly}, as opposed to the 
{\it classical} 
values $a$ arbitrary, $k=0$. We thus feel that the non-existence
of a full and symplectic polarization, for some values of the 
(pseudo-)extension parameters, 
is more a characteristic of the anomaly phenomenon than the appearance 
of deformed terms in the Lie algebra. 


\subsection{Higher-order Polarizations}
\label{higher-order}
In general, to tackle  situations like those mentioned above, it is necessary 
to generalize the notion of polarization. Let us consider the universal 
enveloping algebra of left-invariant vector fields, ${\cal U}\tilde{\cal G}^L$. 
We define a {\bf higher-order polarization} ${\cal P}^{HO}$ as a maximal 
subalgebra of ${\cal U}\tilde{\cal G}^L$ with no intersection with the
abelian subalgebra of powers of $\Xi$. With this definition a higher-order 
polarization contains the maximal number of conditions compatible 
with the equivariance condition of the wave functions and with the action 
of the physical operators (right-invariant vector fields).

We notice that now the vector space of functions annihilated by a
higher-order polarization is not, in general, a ring of 
functions and therefore there is no corresponding foliation; that is, 
they cannot be characterized by saying that they are constant along 
submanifolds. If this were the case, it would mean that the 
higher-order polarization was the enveloping algebra of a first-order 
polarization and, accordingly, we could consider the submanifolds 
associated with this polarization. In this sense the concept of 
higher-order polarization generalizes and may replace that of first-order 
polarization.

We arrive at the formulation of our main general proposition, which had been
proved only for the particular case of the Virasoro group \cite{Virasoro}.
We should stress that the introduction of the (left and right) enveloping 
algebras implies the use of pseudo-differential operators. However, there can 
be non-trivial operators acting on the Hilbert space of wave functions which 
are not pseudo-differential, and therefore are not contained in the enveloping 
algebras. This imply that higher-order polarizations will provide only a 
Hilbert space on which any differential or pseudo-differential operator 
commuting with the representation is a multiple of the identity, but possibly 
containing invariant subspaces which are undistinguishable under the action 
of all the operators in the group. Such a representation will be called 
{\it quasi-irreducible}.

\ni {\bf  Proposition}: {\it Let ${\cal P}^{HO}$ be a higher-order polarization
on \Gt. On subspaces characterized by}
\begin{equation}
L_{\Xi}\psi=i\psi\,,\;\;A.\psi=0\;\; \forall  A\in {\cal P}^{HO} 
\label{theorem}\,,
\end{equation}

\ni {\it all the right-invariant vector fields} $\XR$ 
{\it act quasi-irreducibly. Therefore 
the present quantization procedure 
gives rise to a quasi-irreducible representation of the group} 
$\tilde{G}$, {\it provided it is connected and simply connected}.

\ni The proof uses the fact that in a canonical chart at the identity,
the group law $g''=g'*g$ proves to be a {\it formal group law} \cite{Serre} 
and any
translation (composition) on the group admits a unique formal power-series 
expansion, as well as the following 

\ni {\bf Lemma}: {\it If $\hat{O}$ is a pseudo-differential operator acting 
on the representation
space in such a way that $[\hat{O},\XR]=0$ for any right-invariant
vector field,
then the operator $\hat{O}$ has necessarily the form $\sum_{n=1}^{\infty}
a_{i_1i_2...i_n}\XL_{i_1}\XL_{i_2}...\XL_{i_n}, \;i_k=1,...,dim \tilde{G}$,
where the coefficients $a_{i_1i_2...i_n}$ are constants}.

\ni The Lemma is a direct consequence of the triviality of the tangent
bundle to any Lie group. In fact, any first order operator on the group
can be written as $\tilde{X}=a^i\XL_i$, where $\{a^i\}$ are arbitrary functions on the
group and $\{\XL_i\}$ a basis for the free module of vector fields. Since
right- and left-invariant vector fields commute on any Lie group, the 
condition $[\tilde{X},\XR_i]=0$ necessarily implies $a^i=const,\;\forall i$. The
same condition is obtained for a higher-order operator, which can be
written as $\sum_{n=1}^{\infty}a_{i_1i_2...i_n}\XL_{i_1}\XL_{i_2}...\XL_{i_n}$.

\ni {\it  Proof of the Proposition}: let us assume that the set of solutions of 
the equations  (\ref{theorem}) carries a reducible (and not quasi-irreducible) representation 
of the Lie algebra of $\tilde{G}$ realized by an infinitesimal left action of the group.
Therefore, and according to Schur's lemma in it inverse form (see e.g. the
classical book by Wigner \cite{Wigner}), a non-trivial (not a multiple of
the identity) pseudo-differential operator $\hat{O}$ must exist, at least, 
which commutes with 
the representation. The previous Lemma then requires the operator $\hat{O}$ 
to have the expression given above, and the fact that this operator preserves
(it commutes with the representation) the solution space of the 
polarization ${\cal P}^{HO}$ just states its 
compatibility with the operators in ${\cal P}^{HO}$, i.e. it closes an algebra
with the elements in the polarization, which, in addition, satisfies the
conditions for a higher-order polarization.
In fact, if $A$ is any operator in ${\cal P}^{HO}$, then 
\begin{equation}
[\hat{O},A]\psi_{\alpha}=\hat{O}A\psi_{\alpha}-
                        \sum O_{\alpha\beta}A\psi_{\beta}=0\,,
\end{equation}

\ni so that, the right hand side of this equation can only be replaced, 
at most, by another element of the polarization $A'$. In particular, the 
commutator $[\hat{O},A]$ will never be proportional to the vertical 
generator $\Xi$. However, this is precisely the 
condition for an element in the left enveloping algebra to enter the 
(higher-order) polarization. Thus, the existence of such a non-trivial 
operators would imply that ${\cal P}^{HO}$ was not maximal.

In the case of infinite-dimensional representations, the validity of  
Schur's lemma requires the unitarity of the representation, a fact that
will restrict, in general, the values of the central charges, as mentioned
above.

The definition of higher-order polarization given above is quite general. In 
all studied examples higher-order polarizations adopt a more definite 
structure closely related to given first-order (non-full and/or 
non-symplectic) ones.
According to the until now studied cases, higher-order polarizations can be 
given a more operative definition: {\it A higher-order polarization is a 
maximal subalgebra of ${\cal U}\tilde{\cal G}^L$ the ``vector 
field content'' of which is a first order polarization}. 
By ``vector field content'' of a subalgebra ${\cal A}$ of  
${\cal U}\tilde{\cal G}^L$ we mean the following: 
Let $V({\cal A})$ be the vector space of 
complex functions on \Gt defined by
\begin{equation}
V({\cal A})=\{f\in {\cal F}_C(\tilde{G})\; /\; A\cdot f=0, \forall A\in {\cal A}\} \,.
\end{equation}

\ni Now we generate a ring $R({\cal A})$ by taking any function of elements 
of $V({\cal A})$. With $R({\cal A})$ we associate the set of left-invariant 
vector fields defined by 
\begin{equation}
L_{\tilde{X}^L} h=0\,, \;\;  \forall h\in R({\cal A})\,.
\end{equation}

\ni This set of left-invariant vector fields is a Lie subalgebra of 
$\tilde{{\cal G}}^L$ and defines the vector field content of ${\cal A}$, which
proves to be a first-order polarization. 

A simple example suggesting the generalization of the concept of higher-order 
polarization corresponds to the non-irreducible representation associated with 
the non-symplectic polarization (\ref{non-symplectic}) of the Sch\"rodinger 
group (see above). This polarization cannot be further reduced by enlarging 
(\ref{non-symplectic}) to a higher-order polarization ${\cal P}^{HO}$. A full 
reduction requires the inclusion in ${\cal P}^{HO}$ of the parity operator 
commuting with the representation. The generalization of the concept of 
higher-order polarization so as to include non-pseudo-differential operators, 
as well as a constructive characterization of those operators deserves a 
separate study.


In order to throw some light on the structure of higher-order polarizations,
let us consider the set ${\cal S}$ of all possible higher-order polarizations 
on \Gt. This is a quite large set, so let us try to find some relations on it.

Suppose that $\PHO$ and $\PHO{}'$ are two different higher-order polarizations
on \Gt, but leading to unitarily equivalent representations of \Gt, that is,
\begin{eqnarray}
\PHO &\Rightarrow& U(g)\,\, \hbox{acting on the Hilbert space}\,\,\H\,, \nn \\
\PHO{}' &\Rightarrow& U'(g)\,\, \hbox{acting on the Hilbert space}\,\,\H'\,,\nn 
\end{eqnarray}
in such a way that there exists a unitary operator $V:\H\rightarrow \H'$, with 
$U'=V U V^{-1}$. Then, it is clear that any element $A'\in \PHO{}'$ can be
obtained as $A'=VAV^{-1}$, for some $A\in \PHO$. Obviously, this introduces
an equivalence relation in $\cal S$, and a partition of it in equivalence 
classes $[\PHO]$, where all elements in $[\PHO]$ are related through unitary
transformations. 

Let ${\cal P}^1$ be a fisrt-order polarization (either horizontal or 
non-horizontal, but always full and symplectic), such that 
it leads to a (quasi-) irreducible representation of \Gt (this means that no 
anomaly is present). Define ${\cal UP}^1$ as the enveloping algebra of 
${\cal P}^1$ by the whole ${\cal U}\tilde{\cal G}^L$, that is, 
\begin{equation}
{\cal UP}^1 \equiv \left\{ A=\sum_{k=1}^m A_k \XL_{k}\,,\,\, 
   \hbox{such that}\,\,A_k \in {\cal U}\tilde{\cal G}^L \right\}\,,
\end{equation}
where $\{\XL_k\}_{k=1}^m$ is a basis for ${\cal P}^1$, and 
$m = {\rm dim}\,{\cal P}^1$. Then it is easy to check that ${\cal UP}^1$ is
a higher-order polarization. In fact, it does not contain the vertical 
generator $\Xi$ (nor its powers) because ${\cal P}^1$ does not contain it either, and 
${\cal UP}^1$ it is maximal, otherwise the representation obtained with ${\cal P}^1$ would
not be (quasi)-irreducible. 

Consider the class $[{\cal UP}^1]$ of all
higher-order polarizations in $\cal S$ unitarily equivalent to ${\cal UP}^1$.
For certain groups, as for instance, (finite) semisimple and solvable groups 
\cite{Kirillov}, it is always possible to find an ``admissible" subalgebra associated 
with any quantization 1-form $\Theta$ or, in other words, it is always possible 
to find a full and symplectic first-order polarization. This means that for
these groups ${\cal S} = \cup_{\alpha\in I} [{\cal UP}^1]_{\alpha}$, where the
set $I$ parameterizes the (quasi-)irreducible representations of \Gt (which 
are associated to coadjoint orbits).

Note that, from the very construction of ${\cal UP}^1$, it admits a basis
(of vector fields in this case), finite for finite-dimensional groups, and
this structure is translated to all ${\cal P}^{HO} \in [{\cal UP}^1]$, the
basis being of the form $\{V\XL_{k}V^{-1}\}_{k=1}^m$. For this kind of higher-order 
polarizations, this allows us to define the {\it dimension} of a 
higher-order polarization as the dimension of the first-order polarization
to which it is equivalent. Although we do not give a proof, it is reasonable
that all higher-order polarizations, even if they are not equivalent to a 
first-order one, admit a finite basis (for finite-dimensional groups), and 
this is what happens, for instance, with the anomalous Schr\"odinger and 
Virasoro groups (this last example admits an infinite basis, since the group 
is infinite dimensional, but it is countable). 

Finally, let us comment on the structure of the wave functions that are solutions of a 
higher-order polarization. For the case of a first-order polarization, the
solutions of $\XL \Psi =0\,,\forall\XL\in {\cal P}^1$ have always the form
\begin{equation}
\Psi =\z e^{i\chi(g)}\Phi\,,
\end{equation}
where $\chi(g)$ is a real function on $G$. The functions $\Phi$ are defined on 
a Lagrangian submanifold $S$ of \Gt (with respect to $d\Theta$). We can think of 
the unitary operator $e^{i\chi(g)}$ as relating the representation in terms of 
the right-invariant vectorfields $\XR$ acting on the subspace of $L^2(\Gtm)$ 
of polarized functions, to the representation in terms of the (first-order) 
differential operators $e^{-i\chi(g)} \XR e^{i\chi(g)}$ acting on the Hilbert 
space $L^2(S)$. Note that the unitary operator $e^{i\chi(g)}$ is diagonal, and this is 
related to the fact that $S$ is a Lagrangian submanifold
of \Gt.

However, for higher-order polarizations, the wave functions, solutions of
$A\Psi = 0\,,\forall A\in {\cal P}^{HO}$, do not define a foliation in \Gt. But
they can be written, at least formally, as:
\begin{equation}
\Psi=\z e^{i\hat{O}}\Phi\,,
\end{equation}
where $\hat{O}$ is a pseudo-differential operator on \Gt. 
The reason is that, in general, the polarization equations have the form, or 
can be taken (formally) to the form:
\begin{equation}
i\parcial{g^s}\Psi = \hat{O}_s \Psi\,,\,\,\,s=1,\ldots,k\,,
\end{equation}
where $k$ is the dimension of ${\cal P}^{HO}$. The function $\Phi$ belongs to $L^2(S)$, 
although $S$ is not a Lagrangian submanifold of \Gt.

As in the case of first-order polarizations, we can think of the unitary 
operator $e^{i\hat{O}}$ as stating the equivalence of the representation
defined by the right-invariant vector fields $\XR$, acting on the subspace of 
$L^2(\Gtm)$ defined by the polarized wave functions, and the representation defined
by the (higher-order) differential operators $e^{-i\hat{O}} \XR e^{i\hat{O}}$,
acting on $L^2(S)$.


To see how a higher-order polarization operates in practice, we shall 
consider the examples of the anomalous Schr\"odinger and Virasoro groups. 



\subsection{The Schr\"odinger group}
\label{Chorri}

To illustrate the Lie algebra structure of an anomalous group, let us 
consider the example of the Schr\"odinger group in one dimension. 
This group, or rather, the 
complete Weyl-Symplectic group, was considered in Ref. \cite{Kirillov} as an 
example of a group not possessing an ``admissible" subalgebra. It is the 
semidirect action of the $SL(2,R)$ group on the H-W group (when considered in 
$n$ dimensions, rotations should also be included) including as subgroups the 
symmetry group of the free particle, the Galilei group, as well as the 
symmetry group of the ordinary harmonic oscillator and the ``repulsive'' 
harmonic oscillator (with imaginary frequency), usually known as Newton groups 
\cite{Niederer}. From the mathematical point of view, it can be obtained from 
the Galilei group (or from either of the Newton groups) by replacing the time 
subgroup with the three-parameter group $SL(2,R)$. In fact, those kinematical 
subgroups are associated with different choices of a Hamiltonian inside 
$SL(2,R)$.


In order to perform a global-coordinate treatment of the problem, we shall 
start by considering matrices $S\in GL(2,R)$ instead of $SL(2,R)$, and the 
condition for these matrices to belong to $SL(2,R)$ will appear naturally.
A group law for the Schr\"odinger group can be written as:
\begin{eqnarray}
\x{}\ ''&=& |S|^{1/2} S^{-1} \x\ ' + \x \nn \\
S'' &=& S' S \\
\z''&=&\z'\z \exp\frac{im\w}{2\hbar}\left[
 \frac{-Ax_2'x_1-Bx_2'x_2+Cx_1'x_1+Dx_1'x_2}{|S|^{1/2}}\right]\,,\nn
\end{eqnarray}

\ni where $\x=(x_1,x_2) \in R^2$, 
$S=\left(\begin{array}{lr} A&B\\C&D\end{array}\right)\in GL(2,R)$,  
$|S|\equiv AD-BC$ and $\frac{m\w}{\hbar}$ is a constant 
parametrizing the central extensions of the H-W group (we write it in this 
form for later convenience). The factor $|S|^{1/2}$ in the semidirect 
action of $GL(2,R)$ is needed in order to have a proper central extension. 
For the moment, we shall assign no dimensions to $x_1$ and $x_2$, but if we 
want to identify them with $q$ and $v$ of the H-W group in section 
\ref{H-W}, and keep $SL(2,R)$ adimensional, then it should be $x_1=q$ and
$x_2=v/\w$, where $\w$is a constant having the dimensions of a frequency ($T^{-1}$).

{}From the group law, the left-invariant vector fields associated with the 
coordinates $x_1,x_2,A,B,C,D,\z$, 
\begin{eqnarray}
\XL_{x_1} &=& \parcial{x_1} - \frac{m\w}{2\hbar}x_2\Xi \nn\\
\XL_{x_2} &=& \parcial{x_2} + \frac{m\w}{2\hbar}x_1\Xi \nn\\
\XL_{A} &=& A\parcial{A} + C\parcial{C}-\medio x_1\parcial{x_1}
          +\medio x_2\parcial{x_2} \nn \\
\XL_{B} &=& A\parcial{B} + C\parcial{D} - x_2\parcial{x_1} \\
\XL_{C} &=& B\parcial{A} + D\parcial{C} - x_1\parcial{x_2} \nn \\
\XL_{D} &=& B\parcial{B} + D\parcial{D}+\medio x_1\parcial{x_1}
           - \medio x_2\parcial{x_2} \nn \\
\XL_{\z} &=& i\z\parcial{\z}\equiv \Xi\,, \nn 
\end{eqnarray}

\ni as well as the right-invariants ones,
\begin{eqnarray}
\XR_{x_1} &=& |S|^{-1/2}\left[ D\parcial{x_1} - C\parcial{x_2}
              + \frac{m\w}{2\hbar}(Dx_2+Cx_1)\Xi\right] \nn \\
\XR_{x_2} &=& |S|^{-1/2}\left[ A\parcial{x_2} - B\parcial{x_1}
              - \frac{m\w}{2\hbar}(Bx_2 + Ax_1)\Xi\right] \nn \\
\XR_{A} &=& A\parcial{A} + B\parcial{B}  \nn \\
\XR_{B} &=& D\parcial{B} + C\parcial{A}  \\
\XR_{C} &=& A\parcial{C} + B\parcial{D} \nn \\
\XR_{D} &=& C\parcial{C} + D\parcial{D} \nn\\
\XR_{\z}&=& \Xi\,, \nn 
\end{eqnarray}

\ni can be obtained. The commutation relations for the (left) Lie algebra are:
\be
\begin{array}{lcl}
\left[ \XL_{A},\XL_{B} \right] &=& \XL_{B} \\
\left[ \XL_{A},\XL_{C} \right] &=& -\XL_{C} \\
\left[ \XL_{A},\XL_{D} \right] &=& 0 \\
\left[ \XL_{B},\XL_{C} \right] &=& \XL_{A}-\XL_{D} \\
\left[ \XL_{B},\XL_{D} \right] &=& \XL_{B} \\
\left[ \XL_{C},\XL_{D} \right] &=& -\XL_{C} \\
\left[ \XL_{x_1},\XL_{x_2} \right] &=& \frac{m\w}{\hbar}\Xi \\
\left[ \XL_{A},\XL_{x_1} \right] &=& \medio\XL_{x_1} 
\end{array}
\begin{array}{lcl}
\left[ \XL_{A},\XL_{x_2} \right] &=& -\medio\XL_{x_2} \\
\left[ \XL_{B},\XL_{x_1} \right] &=& 0 \\
\left[ \XL_{B},\XL_{x_2} \right] &=& \XL_{x_1} \\
\left[ \XL_{C},\XL_{x_1} \right] &=& \XL_{x_2} \\
\left[ \XL_{C},\XL_{x_2} \right] &=& 0\\
\left[ \XL_{D},\XL_{x_1} \right] &=& -\medio\XL_{x_1} \\
\left[ \XL_{D},\XL_{x_2} \right] &=& \medio\XL_{x_2}\,. 
\end{array}
\ee

{}From these commutation relations we see that two linear combinations of
vector fields can be introduced, $\XL_{A}-\XL_{D}$ and $\XL_{A}+\XL_{D}$ (the same
for the right-invariant vector fields), in such a way that $\XL_{A}+\XL_{D}$ 
is a central generator, which is also horizontal, and, therefore,
it is a gauge generator (see \cite{gauge}, and also subsection \ref{SU(2)}, where 
something analogous happens for the case of the $SU(2)$ group). 
In fact, it coincides with its right version, as is always the case for a 
central generator.

The quantization 1-form $\Theta$ is: 
\begin{eqnarray}
\Theta &=& \frac{m\w}{2\hbar}(x_2dx_1-x_1dx_2 + 
   \frac{1}{|S|}\left[Cx_1^2dA + Dx_2^2dB \right. \nn \\
 & & \left. -Ax_1^2dC-Bx_2^2dD+x_1x_2d|S|\right]) + \frac{d\z}{i\z} \,,
\end{eqnarray}

\ni and the characteristic subalgebra has the form: 
\begin{equation}
{\cal G}_{\Theta} = <\XL_{A}+\XL_{D}, \XL_{A}-\XL_{D}, \XL_{B},\XL_{C} >\,.
\end{equation}

The Noether invariants, as introduced in Sec. \ref{GAQ}, turn out to be:
\begin{equation}
\begin{array}{lcl}
F_A & =& \frac{m\w}{2\hbar} x_1 x_2 \\
F_B & =& \frac{m\w}{2\hbar} x_2^2 \\
F_C &=& -\frac{m\w}{2\hbar} x_1^2 
\end{array}
\qquad
\begin{array}{lcl}
F_D &=& -\frac{m\w}{2\hbar} x_1x_2 \\
F_{x_1} &=& \frac{m\w}{\hbar} x_2 \\
F_{x_2} &=& - \frac{m\w}{\hbar} x_1 \,,
\end{array}
\end{equation}  

\ni revealing that the Noether invariant associated with the 
right-invariant vector fields whose left counterpart lie in the 
characteristic subalgebra can be expressed in terms of the basic Noether 
invariants $F_{x_1}$ and $F_{x_2}$:
\begin{equation}
\begin{array}{lcl}
F_A + F_D & =& 0 \\
F_A - F_D & =& -\frac{\hbar}{m\w} F_{x_1} F_{x_2}
\end{array}
\qquad
\begin{array}{lcl}
F_B &=& \frac{\hbar}{2m\w} F_{x_2}^2\\
F_C &=& -\frac{\hbar}{2m\w} F_{x_1}^2 \,,
\end{array} \label{Noetherel}
\end{equation}  

\ni Note also that the Noether invariant associated with the 
generator $\XR_A+\XR_D$ is zero, as corresponds to a gauge generator
\cite{gauge}.
These relations between Noether invariants lead to a phase space of the 
classical theory of dimension two (i.e. only one degree of freedom). We have 
obtained this result because the $SL(2,R)$ subgroup is represented trivially, 
that is, the two-cocycle we have chosen for the Schr\"odinger group lies in the 
trivial pseudo-cohomology class, associated with the zero-dimensional 
coadjoint orbit of $SL(2,R)$. Therefore, only the 2-dimensional coadjoint 
orbit of the H-W group is taken into account. See \cite{Perroud,symplin} for 
the study of the Schr\"odinger group considering all coadjoint orbits of
$SL(2,R)$.

As we have seen, the Lie-algebra two-cocycle $\Sigma$ contains the entire 
$SL(2,R)$ subalgebra (together with the gauge generator) in its kernel, which, 
according to the general scheme, should enter any full and symplectic 
polarization. Unfortunately such a polarization does not exist, and this can 
be traced back to the fact
that the partial complex structure $J=\theta^L{}^{x_1}\otimes \XL_{x_2} - 
\theta^L{}^{x_2}\otimes\XL_{x_1}$ is not preserved by its kernel (the $SL(2,R)$ 
subalgebra), implying that we cannot relate it to a complex structure on 
the classical phase-space. We can only find a non-symplectic, full 
polarization, 
\begin{equation}
{\cal P}_C=<\XL_A+\XL_D, \XL_A-\XL_D,\XL_B,\XL_C>\,, \label{non-symplectic}
\end{equation}
 
\ni and a series of non-full, but symplectic polarizations. The latters are 
obtained by adding, to the polarizations given in section \ref{H-W} for the H-W 
group, the gauge generator $\XL_{A}+\XL_{D}$ and two (properly selected) 
generators out of the $SL(2,R)$ subgroup.

Quantizing with these non-full polarizations  
results in a breakdown of the naively expected relations between the 
operators $\XR_A-\XR_D,\XR_B,\XR_C$ and the basic ones $\XR_{x_1},\XR_{x_2}$, 
as suggested by the relations between Noether invariants (\ref{Noetherel}).

Furthermore, quantizing with the non-symplectic polarization 
(\ref{non-symplectic}) leads to an unconventional representation in which the 
wave functions depend on both $x_1$ and $x_2$ variables, and contain two 
irreducible components (see \cite{Hove}). Also, neither of the operators
$\XR_A-\XR_D,\XR_B$, nor $\XR_C$, can be expressed in terms of the basic ones.
 
In all these cases, the operators of $SL(2,R)$ behave as if they had 
symplectic content; that is, as if they were associated with a new degree of
freedom (see \cite{Perroud} and \cite{symplin} for different discussions about
the meaning of this degree of freedom). Thus, there is a breakdown of the
correspondence between (classical) coadjoint orbits and (quantum) Hilbert
spaces.

 Nevertheless, it is still possible to obtain a quantum theory associated with
the classical theory given by the Noether invariants (\ref{Noetherel}). And 
this is given by means of a higher-order polarization of the form:
\begin{eqnarray}
{\cal P}^{HO}&=&<\XL_{A}+\XL_{D},\XL_{A}-\XL_{D}-
   \frac{i\hbar}{2m\w}\left(\XL_{x_1}\XL_{x_2}+\XL_{x_2}\XL_{x_1}\right), \nn\\
& & \XL_{B}+\frac{i\hbar}{2m\w}\left(\XL_{x_1}\right)^2, 
 \XL_{C}-\frac{i\hbar}{2m\w}\left(\XL_{x_2}\right)^2> \,\cup\,  {\cal P}^1 \,,
\label{HOPol}
\end{eqnarray}

\ni where ${\cal P}^1$ is any of the (first-order) polarizations for the
H-W group given in section \ref{H-W} (taking there k=2, so that we can forget 
about the generators $\XL_{\a}$). Note that the higher-order elements of the
polarizations correspond (apart from normal order) to making the substitution
$F_g^i \rightarrow -i \XL_{g^i}$ in the relations among Noether invariants
(\ref{Noetherel}). It should be stressed that the normal order is 
automatically fixed, precisely by the requirement for ${\cal P}^{HO}$ of being a higher-order
polarization. In more general situations, there will be different possibilities in the
normal order, in correspondence with different choices of 
higher-order polarizations. 

For each choice of ${\cal P}^1$, it is more convenient to express the 
higher-order terms in a different basis, to obtain the representation in the
adequate parameters for the description of the free particle, the 
harmonic oscillator or the repulsive oscillator. Thus, the different 
higher-order polarizations could be:
\begin{eqnarray}
{\cal P}^{HO}_v &=& <\XL_{A}+\XL_{D},\,\XL_{A}-\XL_{D}-
   \frac{i\hbar}{2m\w}\left(\XL_{x_1}\XL_{x_2}+\XL_{x_2}\XL_{x_1}\right), \nn\\
& & \XL_{B}+\frac{i\hbar}{2m\w}\left(\XL_{x_1}\right)^2,\, 
 \XL_{C}-\frac{i\hbar}{2m\w}\left(\XL_{x_2}\right)^2,\, \XL_{x_1} > \,,\nn \\
& & \nn \\
{\cal P}^{HO}_c &=& <\XL_{A}+\XL_{D},\,(\XL_{A}-\XL_{D}) +i(\XL_B-\XL_C) -
       \frac{\hbar}{m\w} \left(\XL_c\right)^2, \nn\\ 
& &  (\XL_{A}-\XL_{D}) - i(\XL_B-\XL_C) +
       \frac{\hbar}{m\w} \left(\XL_{c^*}\right)^2, \nn\\
& &   \XL_B-\XL_C + \frac{i\hbar}{2m\w}\left(\XL_{c}\XL_{c^*}+
        \XL_{c^*}\XL_{c}\right), \,\XL_{c^*} >, \\
& & \nn \\
{\cal P}^{HO}_{\rho_{-}} &=& <\XL_{A}+\XL_{D},\,(\XL_{A}-\XL_{D}) + 
  (\XL_B-\XL_C) + \frac{i\hbar}{m\w} \left(\XL_{\rho_{-}}\right)^2, \nn \\
 & & (\XL_{A}-\XL_{D}) - (\XL_B-\XL_C) - 
       \frac{i\hbar}{m\w} \left(\XL_{\rho_{+}}\right)^2, \nn \\
& &   \XL_B+\XL_C + \frac{i\hbar}{2m\w}\left(\XL_{\rho_{+}}\XL_{\rho_{-}}+
        \XL_{\rho_{-}}\XL_{\rho_{+}}\right), \,\XL_{\rho_{+}} >\,. \nn 
\end{eqnarray}

It should be stressed that all these higher-order polarizations are equivalent, 
in the sense that they lead to (unitarily) equivalent representations of the 
Schr\"odinger group. However, one can select certain uniparametric subgroups
of $SL(2,R)$ to assign a dynamics to the H-W group, and obtain in this 
way three differents subgroups of the Schr\"odinger group: the Galilei group of
the free non-relativistic particle, the harmonic oscillator group, and the
repulsive harmonic oscillator group. The three different higher-order 
polarizations ${\cal P}^{HO}_{v}$, ${\cal P}^{HO}_{c}$ and 
${\cal P}^{HO}_{\rho_{-}}$ are adequate for the description of each one of these
systems because they incorporate, explicitely, the corresponding Schr\"odinger
equation, as a polarization equation, which turns out to lead to a 
first-order differential equation.

Let us obtain the representation associated with ${\cal P}_v^{HO}$. The
polarization equations are:
\begin{eqnarray}
\left(\XL_{A}+\XL_{D}\right)\Psi&=&0 \nn\\
\XL_{B}\Psi&=&0 \nn\\
\left(\XL_{A}-\XL_{D}\right)\Psi&=&-\medio\Psi \\
\XL_{x_1}\Psi&=& 0 \nn \\
\XL_{C}\Psi &=& \frac{i\hbar}{2m\w}\left(\XL_{x_2}\right)^2\Psi\,, \nn
\end{eqnarray}

The first of these equations has as solutions those complex wave functions 
on the group $GL(2,R)$, which are defined on $SL(2,R)$, as expected. 
Therefore, the solutions of this equation have the form (once the usual 
$U(1)$-equivariant condition is imposed):
\be
\Psi=\z\Phi(a,b,c,d,x_1,x_2)\,,
\ee

\ni where $a\equiv \frac{A}{\sqrt{AD-BC}},b\equiv \frac{B}{\sqrt{AD-BC}},
c\equiv \frac{C}{\sqrt{AD-BC}}$ and $d\equiv \frac{D}{\sqrt{AD-BC}}$, with
$ad-bc=1$.

To proceed further in solving the polarization equations, it is convenient to 
introduce local charts on $SL(2,R)$. We choose them ads the ones defined 
by $a\neq 0$ and $c\neq 0$, respectively\footnote{Certainly they really 
correspond to four contractible charts: $a>0,a<0$ and $c<0,c>0$, but the 
transition functions between each pair of these charts are trivial, so that we
shall consider them as only one chart.}. The first chart contains the identity 
element $I_2$ of $SL(2,R)$, and the second contains 
$J\equiv\left( \begin{array}{cc}0&1\\-1&0\end{array}\right)$.

The solutions to the the polarization equations are given by:
\begin{itemize}
\item For $a\neq 0$:
\be
\Psi=\z a^{-1/2} e^{\frac{im\w}{2\hbar}xy}\chi(\tau,y)\,, \label{polarizadas1}
\ee

\ni where $x\equiv a(x_1+\frac{b}{a}x_2),\ y\equiv x_2/a$ and 
$\tau\equiv \frac{c}{a}$, with $\chi$ satisfying the Schr\"odinger-like 
equation
\be
\frac{\partial\chi}{\partial\tau} = 
\frac{i\hbar}{2m\w}\frac{\partial^2\chi}{\partial y^2}\,. \label{chorrilike}
\ee

\item For $c\neq 0$:
\be
\hPsi=\z c^{-1/2} e^{-\frac{im\w}{2\hbar}\hx\hy}\hchi(\htau,\hy)\,,
\ee

\ni where $\hx\equiv x_2/c,\ \hy\equiv a(x_1-\htau x_2)$ and 
$\htau\equiv \frac{a}{c}$,
with $\hchi$ satisfying the Schr\"odinger-like equation
\be
\frac{\partial\hchi}{\partial\htau} = 
-\frac{i\hbar}{2m\w}\frac{\partial^2\hchi}{\partial \hy^2}\,.
\ee
\end{itemize}

The element $J$ represents a rotation of $\frac{\pi}{2}$ in the plane 
$(x_1,x_2)$, and takes the wave function from one local chart to the 
other\footnote{In fact, up to a factor, $J$ represents the Fourier transform 
passing from the $x_1$ representation to the $x_2$ representation.}. 
Obviously, $J^4=I_2$, but acting with $J$ on 
the wave functions we obtain:
\be
\Psi(J*g)=(-1)^{1/4}\hPsi(g)\,,
\ee

\ni from which the result $\Psi(J^4*g)=-\Psi(g)$ follows; that is, the 
representation obtained for the subgroup $SL(2,R)$ is two-valued. This 
representation is the well-known {\it metaplectic} or {\it spinor 
representation}. The 
metaplectic representation is for $SL(2,R)$ what the $\medio$-spin representation
is for $SO(3)$ (see \cite{Folland} and references therein, and also 
\cite{Kirillov}). 

If now one computes the action of the right-invariant vector fields on these 
wave functions (now we shall forget about the second chart, since we simply 
have to ``Fourier"-transform with $J$ to obtain the results in the other chart), we
obtain:
\begin{eqnarray}
\XR_{x_1}\Psi &=& \z a^{-1/2} e^{\frac{im\w}{2\hbar}xy}\left[ i\frac{m\w}{\hbar}y 
          - \tau\parcial{y}\right]\chi(\tau,y) \nn \\
\XR_{x_2}\Psi &=& \z a^{-1/2} e^{\frac{im\w}{2\hbar}xy}\left[ 
                \parcial{y}\right]\chi(\tau,y) \nn \\
(\XR_{A}-\XR_{D})\Psi &=& \z a^{-1/2} e^{\frac{im\w}{2\hbar}xy}\left[ - y\parcial{y} -\medio
            - \tau\parcial{\tau}\right]\chi(\tau,y)  \\
\XR_{B}\Psi &=& \z a^{-1/2} e^{\frac{im\w}{2\hbar}xy}\left[ i\frac{m\w}{2\hbar}y^2 
              - \tau^2\parcial{\tau} - \tau(y\parcial{y}+\medio)\right]\chi(\tau,y) \nn \\
\XR_{C}\Psi &=& \z a^{-1/2} e^{\frac{im\w}{2\hbar}xy}\left[ 
             \parcial{\tau}\right]\chi(\tau,y)\,. \nn 
\end{eqnarray}

It is easy to check that the full reduction has been achieved, since we can reproduce the
relation between Noether invariants:
\begin{eqnarray}
\XR_B\Psi &=& -i \frac{\hbar}{2m\w} \left(\XR_{x_1}\right)^2 \Psi \nn \\
\XR_C\Psi &=& i \frac{\hbar}{2m\w} \left(\XR_{x_2}\right)^2 \Psi  \\
(\XR_A-\XR_D)\Psi &=& i \frac{\hbar}{2m\w} 
           \left(\XR_{x_1}\XR_{x_2}+\XR_{x_2}\XR_{x_1}\right) \Psi \,.\nn 
\end{eqnarray}

We can give a more compact form to the expressions of the right-invariant vector fields, 
by solving (formally) the Schr\"odinger-like  equation in (\ref{chorrilike}), so that we 
can write $\chi(\tau,y) = e^{i\frac{\hbar}{2m\w}\tau\frac{d^2}{dy^2}}\phi(y)$, where 
now $\phi(y)$ is an arbitrary funtion of $y$, representing the unique degree of freedom of the
system. As indicated in Sec. \ref{higher-order}, we can obtain the expressions for the
action of the group \Gt in $L^2(S)$, where $S$ is the manifold defined by the coordinate
$y$ (in this case $S=R$), performing the unitary transformation mentioned there; that is,
$X^{HO}\equiv e^{-i\hat{O}}\XR e^{i\hat{O}}$, where $e^{i\hat{O}}$ is the unitary operator
\begin{equation}
e^{i\hat{O}} = a^{-1/2} e^{\frac{im\w}{2\hbar}xy}e^{i\frac{\hbar}{2m\w}\tau\frac{d^2}{dy^2}}\,.
\end{equation}

 The expresion of the pseudo-differential operators $X^{HO}$ acting on $L^2(S)$ can be obtained
(with a bit of care, since in general one has to apply the Campbell-Haussdorf formula and the 
relation 
$e^{-A}Be^{A} = B + [B,A] + \frac{1}{2!}[[B,A],A] +\frac{1}{3!}[[[B,A],A],A] + \cdots$), 
proving to be:
\begin{eqnarray}  
X^{HO}_{x_1} &=& i\frac{m\w}{\hbar} y \nn \\
X^{HO}_{x_2} &=& \frac{d}{dy} \nn \\
X^{HO}_{A} - X^{HO}_D &=& -y\frac{d}{dy} -\medio  \\
X^{HO}_{B} &=& i\frac{m\w}{2\hbar} y^2 \nn \\
X^{HO}_{C} &=& i\frac{\hbar}{2m\w} \frac{d^2}{dy^2} \,.\nn 
\end{eqnarray}

Note that the operator $X^{HO}_{A} - X^{HO}_D$ (which is nothing other than the dilation 
operator) appears with the correct ordering, i.e. it is  
symmetrized. This fact is traced back to the form of the higher-order polarization 
${\cal P}_v^{HO}$, which imposes the correct normal form to all the operators, and that is
uniquely determined by the very definition of higher-order polarization. That is, with the
formalism of higher-order polarizations there is no need of imposing symmetrization ``by hand"
to the final operators. 

As commented before, this polarization is adequate for the description of the free 
non-relativistic particle. We only have to restrict ourselves to the subgroup of matrices
$S$ of the form:
\begin{equation}
S=\left(\begin{array}{lr} 1&\w t\\0&1\end{array}\right)
\end{equation}
for which only the generator $\XR_B$, with $B=\w t$, of the $SL(2,R)$ is relevant. The parameter
$\w$, introduced in order to have adimensional matrices, will disappear in the final expressions,
as expected, since in the Galiei group the only extra parameter is the mass $m$ (and $\hbar$). 
The wave functions, restricted to the Galilei subgroup, are 
$\Psi = \z e^{i\frac{m}{\hbar}(q+tv)v}\phi(v)$, where $q\equiv x_1$ and $v\equiv \w x_2$. The 
action of the right-invariant operators, once they are reduced to the space $L^2(S)$ (which in this case,
and for the operators of the Galilei subgroup, is expressed in terms of first-order differential 
operators), is:
\begin{eqnarray}  
\hat{p}\equiv -i\hbar X^{HO}_{x_1} &=& p \nn \\
\hat{q}\equiv -i\frac{\hbar}{\w} X^{HO}_{x_2} &=& -i\hbar\frac{d}{dp} \\
\hat{E}\equiv i\hbar\w X^{HO}_{B} &=& \frac{p^2}{2m} \nn \,,
\end{eqnarray}
where $p\equiv mv$. The Schr\"odinger equation is given by the polarization equation
$\XL_B\Psi =0$, and it simply tells us that $i\hbar\parcial{t}\Psi =\frac{p^2}{2m} \Psi$, 
as usual.

The same precedure can be carried out to obtain the solutions to the polarizations 
${\cal P}_c^{HO}$ and ${\cal P}_{\r_-}^{HO}$, and to obtain the usual expressions for the 
quantum operators, once we
restrict ourselves to the ordinary and repulsive harmonic oscillators, respectively. We shall 
not pursue on it, and send the reader to Ref. \cite{Wolf} for a detailed study on them, 
although there the quantization
of the complete Schr\"odinger group was not carried out --only the different subgroups were 
studied separatedly-- since higher-order polarizations were not available yet. This is the
reason why in \cite{Wolf} the corrective term $\medio$ was introduced ``by hand" in the
expressions of the operators like $\XR_A-\XR_D$, or in the energy operators of the standard
and repulsive oscillators. With the use of higher-order polarizations this is no longer needed,
since all operators appear automatically normally ordered. This is an advantage of the 
technique of higher-order polarizations, which is more natural (we believe) than the 
``metaplectic correction" (see \cite{Woodhouse,Folland}).


If one wishes to obtain the representation in configuration space for 
all of these systems, with higher-order differential Schr\"odinger equations, 
one simply has to choose ${\cal P}_q$ for ${\cal P}^1$, as well as an 
adequate basis for the higher-order terms in such a way that the corresponding
Schr\"odinger equation appears, explicitely, as one of the polarization
equations.

For instance, the higher-order polarization,
\begin{eqnarray}
{\cal P}^{HO}_{\rm Galilei}&=&<\XL_{A}+\XL_{D},\,\XL_{A}-\XL_{D}-
   \frac{i\hbar}{2m\w}\left(\XL_{x_1}\XL_{x_2}+\XL_{x_2}\XL_{x_1}\right), \nn\\
& & \XL_{B}+\frac{i\hbar}{2m\w}\left(\XL_{x_1}\right)^2,\, 
 \XL_{C}-\frac{i\hbar}{2m\w}\left(\XL_{x_2}\right)^2,\, \XL_{x_2}>\,,
\end{eqnarray}

\ni would be appropriate for the description of the free non-relativistic
particle in configuration space, since it incorporates the second-order
Schr\"odinger equation 
$\left[\XL_{B}+\frac{i\hbar}{2m\w}\left(\XL_{x_1}\right)^2\right]\Psi =0$. 
 
In the same way, the higher-order polarization,
\begin{eqnarray}
{\cal P}^{HO}_{\rm Newton}&=&<\XL_{A}+\XL_{D},\,\XL_{A}-\XL_{D}-
   \frac{i\hbar}{2m\w}\left[\XL_{x_1}\XL_{x_2}+\XL_{x_2}\XL_{x_1}\right], \nn\\
& & \XL_{B} - \XL_{C} +\frac{i\hbar}{2m\w}\left[ \left(\XL_{x_1}\right)^2 + 
\left(\XL_{x_2}\right)^2\right],\nn \\ 
& & \XL_{B}+\XL_{C}+\frac{i\hbar}{2m\w}\left[\left(\XL_{x_1}\right)^2 - 
\left(\XL_{x_2}\right)^2\right],\, \XL_{x_2}>\,,
\end{eqnarray}

\ni is adequate for the descrition of the Newton groups, i.e the attractive 
and the repulsive harmonic oscillators, since it incorporates both 
Schr\"odinger equations, 
$\left\{\XL_{B} - \XL_{C} +\frac{i\hbar}{2m\w}\left[ \left(\XL_{x_1}\right)^2 
+ \left(\XL_{x_2}\right)^2\right]\right\}\Psi =0$, for the attractive harmonic 
oscillator, and  
$\left\{\XL_{B}+\XL_{C}+\frac{i\hbar}{2m\w}\left[\left(\XL_{x_1}\right)^2 - 
\left(\XL_{x_2}\right)^2\right]\right\}\Psi =0$ for the repulsive harmonic 
oscillator.

We shall not solve these polarization equations, see, for instance, \cite{oscilata}
and \cite{Marmo2}, for the description of the harmonic oscillator in configuration 
space.

Finally, we would like to make few comments on the geometric interpretation of the anomaly 
of the Schr\"odinger group. As commented at the beginning of this section, the classical 
description of this group leads to a system with one degree of freedom, associated essentially
to the coadjoint orbits (for each value of the parameter $\frac{m\w}{\hbar}$) of the
Heiserberg-Weyl subgroup. That is, the $SL(2,R)$ subgroup, lying in the kernel of 
the symplectic form, is represented trivially; we are in the trivial (the identity
of $sl(2,R)^*$) coadjoint orbit of $SL(2,R)$. Therefore, the classical description has the 
same features as for any other semidirect product, like for instance, the action 
of spatial rotations on the Galilei or the Poincar\'e groups (without spin).  

However, at the quantum level, meanwhile for the Galilei or Poincar\'e groups first-order 
(full and symplectic) polarizations can be found (see \cite{Position}), that is, they are 
not anomalous groups, for the Schr\"odinger group we have to resort to a higher-order 
polarization to obtain the representation associated with only one degree of freedom. Computing 
the Casimir operator for the $SL(2,R)$ subgroup,
\begin{equation}
\hat{C} \equiv \frac{1}{4}\left(\XR_A - \XR_D\right)^2 
             + \medio\left(\XR_B\XR_C + \XR_C\XR_B\right) \,,
\end{equation}
it is easy to check that the representation obtained contains two 
irreducible representations of $SL(2,R)$, corresponding to Bargmann's indices  
$k=\frac{1}{4}$ and $\frac{3}{4}$, and the Casimir operator turns out to be 
$\hat{C}\Psi = -\frac{3}{16}\Psi$ (the relation between the value $c$ of the Casimir and 
the Bargmann's index $k$ is given by $c=k(k-1)$. 
This means that to construct the quantum representation associated with the trivial 
coadjoint orbit of $SL(2,R)$ one must use two different coadjoint orbits, none of them 
being the trivial one. Thus, the Schr\"odinger group is the simplest example
for which the corresponce between coadjoint orbits and quantum Hilbert spaces is broken. 
Another example is that of the Virasoro group, which will be discussed in next section.

An interesting physical application of this construction can be found in Quantum Optics --in 
two-photon systems \cite{Optica}-- where the action of the generators of the $SL(2,R)$ subgroup 
represent the emission
or absorbtion of two photons simultaneously. The two irreducible representations of 
$SL(2,R)$ with $k=\frac{1}{4}$ and $\frac{3}{4}$ correspond to the set of states with even 
and odd number of photons, respectively, and both sets are needed to construct an irreducible
representation of the whole Schr\"odinger group. 


\subsection{The Virasoro group and String Theory}

Let us comment briefly on the relevant, although less intuitive,
example of the infinite-dimensional Virasoro group. Its Lie algebra can 
be written as
\begin{equation}
\left[\XL_{l_n},\XL_{l_m}\right]=-i(n-m)\XL_{l_{n+m}} 
        -\frac{i}{12}(cn^3-c'n)\Xi\,, 
\end{equation}

\ni where $c$ parametrizes the central extensions and $c'$ the 
pseudo-extensions. For the particular case in which
$\frac{c'}{c}=r^2\,,r\in N, r>1$, the co-adjoint orbits admit no invariant 
K\"ahlerian structure, as was firstly stated by Witten pointing out
a breakdown of Geometric Quantization. 
In the present approach, this case shows up as an algebraic anomaly. In 
fact, the characteristic subalgebra is given by
\begin{equation} 
\GC=<\XL_{l_0},\XL_{l_{-r}},\XL_{l_{+r}}>\,,\label{caracteristica} 
\end{equation}

\ni which is not fully contained in the non-full (but symplectic) 
polarization
\begin{equation} 
{\cal P}^{(r)}=<\XL_{l_{n\leq 0}}>\,.\label{Pr} 
\end{equation}
\ni There also exists a full 
polarization ${\cal P}_C=<\XL_{l_{kr}}>\,,r>1, k=-1,0,1,2,3,...$ which
is not symplectic since none of the symplectic generators
with labels $l_{\pm r'}, r'\neq kr$ are included in the polarization. 
A detailed description of the representations of the Virasoro group can be 
found in \cite{Virasoro} and references therein. 

The situation is formally similar to that found in the Schr\"odinger group.
Now, for particular values of the parameters $c,c'$ or, equivalently, 
$c,h\equiv \frac{c-c'}{24}$ given by the Kac formula \cite{Kac}, 
\bea
h &=& \frac{1}{48}(13-c)(k^2+s^2)\pm(c^2-26c+25)(k^2-s^2) \nn \\
  & & -24ks-2+2c \label{Kac} \\
 & & k,\,s\,\,  {\rm positive\ integers},\,\,ks\leq r\,, \nn  
\eea

\ni the ``quantum values" of the anomaly, 
the representations given by the first order non-full (symplectic) 
polarizations (\ref{Pr}) are reducible since there exist invariant subspaces 
characterized by certain higher-order polarization equations 
\cite{Virasoro}. On these subspaces the operators $<\XR_{l_0},\XR_{l_{-r}},
\XR_{l_{+r}}>$, the right version of the characteristic subalgebra 
(\ref{caracteristica}), can be rewritten in terms of the basic operators
$\XR_{l_{k}},\,k\neq\pm r,0$. There is a (standard) anomaly in the sense
that the classical symplectic manifold is reduced (the Noether invariants 
associated with $<\XR_{l_0},\XR_{l_{-r}},\XR_{l_{+r}}>$ are written in 
terms of the basic ones) for the value $c'=cr^2$ or, equivalently, 
$h=-c(r^2-1)/24$, instead of the values given by (\ref{Kac}).
Note that there is no one-to-one correspondence 
between the values of $c'/c$ characterizing the coadjoint orbits of the 
Virasoro group (the classical values of the anomaly) and the values 
allowed by the Kac formula (the quantum values of the anomaly), a fact 
which must be interpreted as a breakdown of the notion of classical 
limit.

String theory provides another example of anomalous system with characteristics
maybe more similar to the finite-dimensional case. The symmetry of the 
bosonic string consists of an infinite set of harmonic oscillators, the
normal modes, on which the Virasoro group acts by semi-direct product. That is,
\bea
\left[\XL_{l_n},\XL_{l_m}\right]&=&-i(n-m)\XL_{l_{n+m}}-i
           \frac{1}{12}(cn^3+c'n)\delta_{n+m,0}\XL_0 \\
\left[\XL_{l_n},\XL_{\alpha_m^{\mu}}\right]&=&-im\XL_{\alpha_{n+m}^{\mu}} \nn\\
\left[\XL_{\alpha_n^{\mu}},\XL_{\alpha_m^{\nu}}\right]&=&-iam\delta_{n+m}\delta^{\mu\nu}\XL_0\;,\label{extension}
\eea

The anomaly is clearly discovered when considering the specific values
$c=0=c'$ of the extension parameters, corresponding to the classical symmetry.
For these values the characteristic subalgebra is
\be
<\XL_{\alpha_0^{\mu}},\,\XL_{l_n}>\,,n\in Z
\ee
\ni and the only allowed polarization, 
\be
{\cal P}=<\XL_{\alpha_{n\leq 0}^{\mu}},\,\XL_{l_{n\leq 0}}>
\ee
\ni is non-full, excluding the Virasoro generators with positive index, 
which also are in the characteristic subalgebra. Quantizing with this 
polarization leads to a Hilbert space containing states obtained from the
vacuum by the action of all the creation operators, i.e. 
$<\XL_{\alpha_{n>0}^{\mu}},\,\XL_{l_{n>0}}>$. However, for the 
``quantum" values of the anomaly, $c=d=c'$, where $d$ is the dimension
of the Minkowski space, the Hilbert space is generated only by states
of the form
\be
\XR_{\alpha_{n_1}^{\mu_1}}...\XR_{\alpha_{n_j}^{\mu_j}}|0>\,,
\ee
\ni and the Virasoro operators are all of them written in terms of the
true basic operators, according to the Sugawara's construction 
\cite{Sugawara}: 
\be
\XR_{l_k}=\frac{1}{2}g_{\mu\nu}:\sum_n\XR_{\alpha_{k-n}^{\mu}}\XR_{\alpha_n^{\nu}}:\,.
\ee
\ni This process is essentially equivalent to the anomalous reduction which
allows the $sl(2,R)$ operators to be written in terms of $\XR_{x_1},\XR_{x_2}$
in the case of the Schr\"odinger group.

\end{document}